\documentclass[journal=jctcce,manuscript=article]{achemso}
\usepackage{cancel}
\usepackage{amsmath}
\usepackage{mathtools}
\usepackage{graphicx}
\usepackage{amssymb}
\usepackage{amsthm}
\usepackage{bm}
\usepackage{dcolumn}
\usepackage{braket}
\usepackage{longtable}
\usepackage{ragged2e}
\usepackage{newtxtext,newtxmath}
\usepackage[version=3]{mhchem} 
\usepackage[T1]{fontenc}       
\usepackage[colorlinks=true,allcolors=blue]{hyperref}
\usepackage[capitalise]{cleveref}

\usepackage[utf8]{inputenc}
\usepackage[version=3]{mhchem} 
\usepackage{amsmath,empheq,mathtools,amssymb}
\usepackage{hyperref}
\usepackage{tikz}
\usetikzlibrary{calc}
\usepackage{siunitx}
\usepackage{enumitem}
\usepackage{braket}
\usepackage{mathtools}
\usepackage[thinc]{esdiff}
\usepackage{physics}
\usepackage{float}

\DeclareUnicodeCharacter{0337}{/}

\allowdisplaybreaks 

\title{Size-Consistent Adiabatic Connection Functionals via Orbital-Based Matrix Interpolation}

\author{Kyle Bystrom}
\email{kbystrom@flatironinstitute.org}
\affiliation{Initiative for Computational Catalysis, Flatiron Institute, New York, New York 10010, USA}
\author{Timothy C. Berkelbach}
\email{t.berkelbach@columbia.edu}
\affiliation{Initiative for Computational Catalysis, Flatiron Institute, New York, New York 10010, USA}
\affiliation{Department of Chemistry, Columbia University, New York, New York 10027, USA}

\SectionNumbersOn

\begin{document}
\begin{singlespace}

\begin{tocentry}

\includegraphics{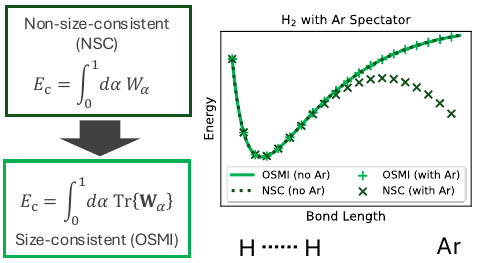}

\end{tocentry}

\begin{abstract}
    We introduce a size-consistent and orbital-invariant formalism for constructing correlation functionals based on the adiabatic connection for density functional theory (DFT).
    By constructing correlation energy matrices for the weak and strong correlation limits in the space of occupied orbitals, our method, which we call orbital-based size-consistent matrix interpolation (OSMI), avoids previous difficulties in the construction of size-consistent adiabatic connection functionals. We design a simple, nonempirical adiabatic connection and a one-parameter strong-interaction limit functional, and we show that the resulting method reproduces the correlation energy of the uniform electron gas over a wide range of densities. When applied to subsets of the GMTKN55 thermochemistry database, OSMI is more accurate on average than MP2 and nonempirical density functionals. Most notably, OSMI provides excellent predictions of the barrier heights we tested, with average errors of less than 2 kcal mol$^{-1}$. Finally, we find that OSMI improves the trade-off between fractional spin and fractional charge errors for bond dissociation curves compared to DFT and MP2. The fact that OSMI provides a good description of molecular systems and the uniform electron gas, while also maintaining low self-interaction error and size-consistency, suggests that it could provide a framework for studying heterogeneous chemical systems.
\end{abstract}

\maketitle

\section{Introduction}

The adiabatic connection formula~\cite{harrisAdiabaticconnectionApproachKohnSham1984,langrethExchangecorrelationEnergyMetallic1975} is a key tool in the design of new correlation functionals for density functional theory (DFT) simulations. One popular use of the adiabatic connection is to approximate the correlation energy by an interpolation between a weak correlation limit of second-order perturbation theory (PT2)---either G\"orling-Levy (GL2)~\cite{gorlingCorrelationenergyFunctionalIts1993,gorlingExactKohnShamScheme1994,gorlingDFTIonizationFormulas1995} or M\o ller-Plesset (MP2)~\cite{mollerNoteApproximationTreatment1934,SzaboOstlund}---and a strong correlation limit approximated by a density functional~\cite{seidlStrictlyCorrelatedElectrons1999}. On their own, finite-order perturbation theory approaches like GL2 and MP2 fail to describe many chemical properties accurately because the perturbation expansion assumes weak electron correlation. By leveraging the adiabatic connection formula to incorporate information about the strong-correlation limit, researchers have designed correlation functionals that provide an improved description of dispersion interactions~\cite{daasNoncovalentInteractionsModels2021,daasCorrectionNoncovalentInteractions2023,daasRegularizedOppositeSpinScaled2023,palosMollerPlessetAdiabatic2025}, molecular reaction energies~\cite{fabianoInteractionStrengthInterpolationMethod2016,constantinAdiabaticconnectionInterpolationModel2024,constantinNonempiricalAdiabaticConnection2025,khanSCANBasedNonlinear2025}, metallic clusters~\cite{giarrussoAssessmentInteractionstrengthInterpolation2018}, and even classic examples of static correlation like hydrogen molecule dissociation~\cite{constantinAdiabaticconnectionInterpolationModel2024,khanSCANBasedNonlinear2025,constantinNonempiricalAdiabaticConnection2025,smigaSelfConsistentImplementationKohn2022}.

However, as discussed in Section~\ref{sec:theory_acxc} below, if the adiabatic connection interpolation is performed over the total weak and strong correlation energies, the correlation model is not size-consistent~\cite{cohenAssessmentFormalProperties2007,mirtschinkEnergyDensitiesStrongInteraction2012,vuckovicRestoringSizeConsistency2018}. A system consisting of two non-interacting subsystems should have an energy equal to the sum of the energies of the two subsystems, but this is not the case for adiabatic connection functionals. In this work, we solve this problem by performing the adiabatic connection interpolation not over scalar correlation energies, but over ``energy matrices'' in the space of occupied orbitals. This results in a method that is naturally both size-consistent and orbital-invariant without any fragment-based energy corrections~\cite{vuckovicRestoringSizeConsistency2018}. This method can be described as an adiabatic connection second-order perturbation theory (ACPT2) with orbital-based size-consistent matrix interpolation (OSMI), and so we abbreviate the method OSMI-ACPT2. (We use the Kohn-Sham adiabatic connection in this work, and so this is a GL2-based method, but the basic technique will be the same for the M\o ller-Plesset adiabatic connection~\cite{seidlCommunicationStronginteractionLimit2018a}.)

To demonstrate the OSMI-ACPT2 formalism, we introduce a simple adiabatic connection interpolation model and benchmark its performance on some subsets of GMTKN55 molecular data,~\cite{goerigkLookDensityFunctional2017a} the uniform electron gas correlation energy, and molecular dissociation curves. The OSMI-ACPT2 is as good or better than the global, non-size-consistent ACPT2 (hereafter NSC-ACPT2) for all of these tests. OSMI-ACPT2 is competitive with or better than state-of-the-art density functional approximations for barrier heights and self-interaction problems, and has reasonable accuracy for bond dissociation energies and dispersion interactions. OSMI-ACPT2 also outperforms canonical MP2 and the regularized $\kappa$-MP2~\cite{Lee2018} on average.

\section{Theory}

\subsection{Adiabatic Connection Functionals} ~\label{sec:theory_acxc}

The Kohn-Sham adiabatic connection formula states that the exchange-correlation (XC) energy can be expressed as a density functional~\cite{ernzerhofConstructionAdiabaticConnection1996,seidlStrictlyCorrelatedElectrons1999}
\begin{equation}
    E_\text{xc}[n] = \int_0^1 \dd\alpha\, W_\alpha[n]\label{eq:adiabatic_connection}
\end{equation}
where
\begin{equation}
    W_\alpha[n] = \mel{\Psi_\alpha[n]}{\hat{V}_\text{ee}}{\Psi_\alpha[n]} - U[n]
\end{equation}
with $\hat{V}_\text{ee}=\sum_{i<j}|\mathbf{r}_i-\mathbf{r}_j|^{-1}$ being the electron-electron repulsion operator and
\begin{equation}
    U[n] = \frac{1}{2} \iint \dd[3]\mathbf{r} \dd[3]\mathbf{r}' \frac{n(\mathbf{r}) n(\mathbf{r}')}{|\mathbf{r}-\mathbf{r}'|}
\end{equation}
being the Hartree energy accounting for classical electron-electron repulsion. The many-body wavefunction $\ket{\Psi_\alpha[n]}$ minimizes the sum of the kinetic energy and electron-electron repulsion $\expval{\hat{T}+\hat{V}_\text{ee}}$, subject to the constraint that it yields the exact ground-state density $n(\mathbf{r})$.
A similar relationship exists for the M\o ller-Plesset perturbation series~\cite{seidlCommunicationStronginteractionLimit2018a,pernalCorrelationEnergyRandom2018,Daas2020,daasGradientExpansionsLargeCoupling2022,daasMollerPlessetAdiabatic2024}.

The adiabatic connection reformulates the problem of finding the XC energy $E_\text{xc}[n]$ as the problem of finding the interaction strength-dependent $W_\alpha[n]$. Although this sounds like a more complicated problem, some exact conditions, including small and large-$\alpha$ limits, are known. For example, the small-$\alpha$ limit satisfies
\begin{equation}
    \lim_{\alpha\rightarrow 0} W_\alpha[n] = W_0[n] + \alpha W_0'[n] \label{eq:ac_scalar_small} 
\end{equation}
where $W_0[n]=E_\text{x}[n]$ is the exact exchange energy of the Kohn-Sham Slater determinant~\cite{seidlStrictlyCorrelatedElectrons1999}, and the first derivative with respect to $\alpha$ at $\alpha=0$ is $W_0'[n]=2E_\text{c}^\text{PT2}$.
The PT2 energy is
\begin{align}
    E_\text{c}^\text{PT2} &= \frac{1}{4} \sum_{ijab} t_{ij}^{ab} \mel{ij}{}{ab} \label{eq:pt2_energy} \\
    t_{ij}^{ab} &= \frac{\mel{ab}{}{ij}}{\epsilon_i + \epsilon_j - \epsilon_a - \epsilon_b} \label{eq:mp2_diag}
\end{align}
where $i,j$ and $a,b$ are occupied and unoccupied molecular spin-orbitals, $\mel{ab}{}{ij}$ is an antisymmetrized two-electron repulsion integral, and the orbitals and their eigenvalues $\epsilon_i$ are those of the mean-field Hamiltonian (Kohn-Sham or Hartree-Fock). In this work, we evaluate the double-excitation contributions only and neglect singles for all ACPT2 models. Neglecting the single excitations is standard practice for double hybrid DFT~\cite{smigaSelfconsistentDoublehybridDensityfunctional2016} (which is closely related to ACPT2), and the singles contribution is typically two orders of magnitude lower than the doubles contribution.~\cite{grabowskiComparisonSecondorderOrbitaldependent2008,grabowskiInitioDensityFunctional2007,grabowskiOrbitaldependentSecondorderScaledoppositespin2014}

The large-$\alpha$ limit is~\cite{seidlStrictlyCorrelatedElectrons1999,seidlStrictlyCorrelatedElectrons1999b,seidlStronginteractionLimitDensityfunctional1999}
\begin{equation}
    \lim_{\alpha\rightarrow\infty} W_\alpha[n] = W_\infty[n] + \alpha^{-1/2} W_\infty'[n] \label{eq:ac_scalar_large}
\end{equation}
where $W_\infty[n]$ and $W_\infty'[n]$ are finite functionals of the density. Other conditions are known, but the focus of this paper is on the OSMI-ACPT2 formalism, so we refer the reader to previous work on the known properties of $W_\alpha[n]$~\cite{seidlStrictlyCorrelatedElectrons1999}.

The small-$\alpha$ case is a well-studied perturbation expansion, while the large-$\alpha$ case can be reasonably approximated by simple density functionals like the point-charge plus continuum model~\cite{seidlDensityFunctionalsStronginteraction2000a}. Therefore, $W_\alpha[n]$ is typically approximated by an interpolation
\begin{equation}
    W_\alpha[n] = W_\alpha(\mathcal{W}[n]) \label{eq:acks_nsc}
\end{equation}
where we have introduced the vector 
\begin{equation}
    \mathcal{W}[n] = (W_0[n], W_0'[n], W_\infty[n], W_\infty'[n], \dots) \label{eq:wfeatures}
\end{equation}
as the features that are used to construct the adiabatic connection. We only use the four explicitly listed features in this work. The exact form of $W_\alpha(\mathcal{W})$ is unknown but should obey the limits in eqs~\ref{eq:ac_scalar_small} and \ref{eq:ac_scalar_large}, and the values for intermediate $\alpha$ are approximated by a heuristic or semi-empirical interpolation. Early interpolations include the SPL~\cite{seidlStrictlyCorrelatedElectrons1999} and ISI~\cite{seidlSimulationAllOrderDensityFunctional2000} models, and more sophisticated approaches have been developed as well~\cite{constantinAdiabaticconnectionInterpolationModel2024} (SPL does not account for the $W_\infty'[n]$ term). In this work, we focus on the ISI model.

It is apparent from eqs~\ref{eq:ac_scalar_small} and~\ref{eq:ac_scalar_large} that $W_\alpha(\mathcal{W}[n])$ must be nonlinear in all of its features. This causes a size-consistency issue. Suppose that a system AB has two non-interacting subsystems A and B. Then physically, we must have $W_\alpha[n^\text{AB}]=W_\alpha[n^\text{A}]+W_\alpha[n^\text{B}]$. However, for nonlinear interpolations,
\begin{align}
    W_\alpha(\mathcal{W}[n^\text{AB}]) &= W_\alpha(\mathcal{W}[n^\text{A}] + \mathcal{W}[n^\text{B}]) \notag\\
    &\ne W_\alpha(\mathcal{W}[n^\text{A}]) + W_\alpha(\mathcal{W}[n^\text{B}])
\end{align}
(where we have assumed that each component of $\mathcal{W}[n]$ is size-consistent)
and therefore the exchange-correlation energy is not size-consistent.
For a practical example of the importance of size-consistency, see Figure~\ref{fig:sc} below, in which an NSC-ACPT2 model produces significantly different dissociation curves for the hydrogen molecule depending on whether there is an Ar atom 100 \r{A} away (while the OSMI-ACPT2 model introduced in this work produces an equivalent dissocation curve regardless of the presence of Ar).

Two approaches exist in the literature for correcting the size-consistency problem. The first is to perform the interpolation locally in real space~\cite{zhouConstructionExchangecorrelationFunctionals2015,vuckovicExchangeCorrelationFunctionals2016,kooiLocalGlobalInterpolations2018,daasExactMollerPlessetAdiabatic2025}, which requires defining an energy density for each input to the adiabatic connection interpolation, e.g.,
\begin{equation}
    W_0'[n] = \int \dd[3]\mathbf{r}\, w_0'[n](\mathbf{r})
\end{equation}
Then the correlation energy is
\begin{equation}
    E_\text{c}[n] = \int \dd[3]\mathbf{r} \int_0^1\dd\alpha\, w_\alpha(\mathcal{W}[n](\mathbf{r}))
\end{equation}
where
\begin{align}
    \mathcal{W}[n](\mathbf{r})=(&w_0[n](\mathbf{r}), w_0'[n](\mathbf{r}),\notag\\
    &w_\infty[n](\mathbf{r}), w_\infty'[n](\mathbf{r}), \dots)
\end{align}
However, there are two key problems with this approach. First, the calculation of the GL2 term $w_0'(\mathbf{r})$ could potentially be computationally expensive due to the large number of grids required to numerically integrate $w_\alpha(\mathcal{W}[n](\mathbf{r}))$ over real space. Second, the energy densities are not uniquely defined because any choice (or gauge) of $w(\mathbf{r})$ that integrates to $W$ is a legitimate definition of the energy density. However, computing different energy densities in different gauges can lead to unphysical behavior at different points in real-space~\cite{gori-giorgiElectronicZeroPointOscillations2009a}. This problem also occurs in local hybrid functionals with the exchange energy density~\cite{maierLocalHybridFunctionals2019}.
We will also show in Section~\ref{sec:ueg} that local interpolation yields the same unphysical behavior as the non-size-consistent approach for the uniform electron gas.
To the best of our knowledge, the local interpolation approach for ACPT2 has only been implemented for small, simple systems.

Another approach is to decompose a chemical system into distinct fragments and then compute a size-consistency correction term for these fragments~\cite{vuckovicRestoringSizeConsistency2018}. This technique has successfully been applied to a variety of chemical systems~\cite{daasNoncovalentInteractionsModels2021,daasRegularizedOppositeSpinScaled2023,palosMollerPlessetAdiabatic2025}; however, there is not an objective or universal way to define the fragments, and the method fails if a system decomposes into fragments with degenerate ground states~\cite{Vuckovic2020}. It would be preferable to design an intrinsically size-consistent approach that does not depend on identifying molecular fragments or performing the interpolation in real space.

\subsection{The OSMI Method} ~\label{sec:theory_osmi}

To construct our adiabatic connection, we start by computing the matrix elements of the $W_\alpha$ inputs in the space of occupied Kohn-Sham orbitals. For example, the $W_0$ matrix is built from the matrix elements of the exchange operator
\begin{equation}
    \left(\mathbf{W}_0[n]\right)_{ij} = \frac{1}{2} \mel{i}{\hat{K}}{j} \label{eq:w0_eqn}
\end{equation}
where
\begin{equation}
    K(\sigma, \mathbf{r}, \sigma' \mathbf{r}') = -\delta_{\sigma\sigma'} \frac{P(\sigma, \mathbf{r}, \sigma', \mathbf{r}')}{|\mathbf{r}-\mathbf{r}'|}
\end{equation}
is the exchange operator, with $\sigma,\sigma'$ being spin indexes and $\mathbf{r},\mathbf{r}'$ being real-space coordinates, and $P$ is the Kohn-Sham 1-particle density matrix,
\begin{equation}
    P(\sigma, \mathbf{r}, \sigma', \mathbf{r}') = \sum_i \phi_i(\sigma,\mathbf{r}) \phi_i^*(\sigma',\mathbf{r}')
\end{equation}
The $W_\infty$ and $W_\infty'$ limits are density functionals and therefore typically computed by integrating the energy density $w_\infty(\mathbf{r})$. For density functional terms like these, we define
\begin{equation}
    (\mathbf{W}_\text{DF}[n])_{ij} = \int \dd[3]\mathbf{r}\, \phi_i^*(\mathbf{r}) \phi_j(\mathbf{r}) \frac{w_\text{DF}[n](\mathbf{r})}{n(\mathbf{r})} \label{eq:wdf_eqn}
\end{equation}
where the ratio $w_\text{DF}[n](\mathbf{r})/n(\mathbf{r})$ has units of energy per particle and $\int \dd[3]\mathbf{r} w_\text{DF}(\mathbf{r})=W_\text{DF}$.

The most complex term for which the matrix must be computed is the perturbation theory term $W_0'$. We choose for this 
\begin{equation}
    \left(\mathbf{W}_0'[n]\right)_{ij} = \frac{1}{4} \sum_{kab} \left(t_{ik}^{ab} \mel{jk}{}{ab} + t_{jk}^{ab} \mel{ik}{}{ab}\right) \label{eq:w0p_eqn}
\end{equation}
which is a symmetric matrix whose trace is twice the PT2 correlation energy of eq~\ref{eq:pt2_energy}. This choice of $\mathbf{W}_0'$ was inspired by the use of this term in the BW-s2 regularized perturbation theory approach~\cite{Carter-Fenk2023}, which uses eq~\ref{eq:w0p_eqn} to modify the zeroth-order Hamiltonian in second-order perturbation theory so that the correlation energy is non-divergent, size-consistent, and invariant to rotations of the occupied or virtual orbitals.

The significance of all the above matrices is that their traces are the corresponding global correlation energy terms,
\begin{align}
    \Tr{\mathbf{W}_0[n]} &= E_\text{x}^\text{exact}[n] = W_0[n] \label{eq:tr_w0} \\
    \Tr{\mathbf{W}_0'[n]} &= 2E_\text{c}^\text{GL2}[n] = W_0'[n] \label{eq:tr_w0p} \\
    \Tr{\mathbf{W}_\text{DF}[n]} &= \int \dd[3]\mathbf{r}\, w_\text{DF}[n](\mathbf{r}) = W_\text{DF}[n] \label{eq:tr_wdf}
\end{align}

Using these correlation energy matrices, the OSMI-ACPT2 XC energy is
\begin{equation}
    E_\text{xc}^\text{OSMI}[n] = \int_0^1 \dd\alpha \Tr{\mathbf{W}_\alpha(\boldsymbol{\mathcal{W}})} \label{eq:osmi_acmp2_xc}
\end{equation}
where $\boldsymbol{\mathcal{W}}=\left\{\mathbf{W}_0, \mathbf{W}_0', \mathbf{W}_\infty', \mathbf{W}_\infty, ...\right\}$ is a list of matrices in the space of occupied molecular orbitals. The function $\mathbf{W}_\alpha(\boldsymbol{\mathcal{W}})$ is a map $\mathbb{R}^{N_\text{feat}\times N_\text{elec} \times N_\text{elec}} \rightarrow \mathbb{R}^{N_\text{elec} \times N_\text{elec}}$, where $N_\text{feat}$ is the number of features in the adiabatic connection interpolation, and $N_\text{elec}$ is the number of electrons. Analogously to eqs~\ref{eq:ac_scalar_small} and~\ref{eq:ac_scalar_large} for the scalar interpolation, the function $\mathbf{W}_\alpha$ should have the limits
\begin{align}
    \lim_{\alpha\rightarrow 0} \mathbf{W}_\alpha(\boldsymbol{\mathcal{W}}[n]) &= \mathbf{W}_0[n] + \alpha \mathbf{W}_0'[n] \label{eq:ac_matrix_small} \\
    \lim_{\alpha\rightarrow \infty} \mathbf{W}_\alpha(\boldsymbol{\mathcal{W}}[n]) &= \mathbf{W}_\infty[n] + \alpha^{-1/2} \mathbf{W}_\infty'[n] \label{eq:ac_matrix_large}
\end{align}
Because of the trace conditions in eqs~\ref{eq:tr_w0},~\ref{eq:tr_w0p}, and~\ref{eq:tr_wdf}, these limits correspond to the correct scalar energy contributions
\begin{align}
    \lim_{\alpha\rightarrow 0} \Tr{\mathbf{W}_\alpha(\boldsymbol{\mathcal{W}}[n])} &= W_0[n] + \alpha W_0'[n] \\
    \lim_{\alpha\rightarrow \infty} \Tr{\mathbf{W}_\alpha(\boldsymbol{\mathcal{W}}[n])} &= W_\infty[n] + \alpha^{-1/2} W_\infty'[n] 
\end{align}

In the following subsections, we explain why the OSMI-ACPT2 XC energy of eq~\ref{eq:osmi_acmp2_xc} is size-consistent and invariant to rotations of the occupied orbitals and of the virtual orbitals. After that, we introduce a specific adiabatic connection interpolation to demonstrate the method.

\subsection{Size-consistency}

Consider two sybsystems A and B of system AB. If A and B are fully separated and non-interacting, then each molecular orbital is localized either in A or B. (In the case of degenerate states, the set of molecular orbitals is non-unique, but a transformation of the orbitals in which they are all localized in A or B can be chosen.) In this space of orbitals, each matrix in $\boldsymbol{\mathcal{W}}$ is block-diagonal (with the blocks being the A and B orbitals). Any operations combining the matrices in $\boldsymbol{\mathcal{W}}$ (addition, multiplication,  exponentiation, etc.) maintain this block-diagonal character, and therefore
\begin{equation}
    \mathbf{W}_\alpha(\boldsymbol{\mathcal{W}}_\text{AB}) = \mathbf{W}_\alpha(\boldsymbol{\mathcal{W}}_\text{A}) + \mathbf{W}_\alpha(\boldsymbol{\mathcal{W}}_\text{B})
\end{equation}
This is the condition for size-consistency.

\subsection{Orbital-invariance}

For OSMI-ACPT2 to be an orbital-invariant theory, we require that
\begin{equation}
    \left(\mathbf{U}\right)^\top \mathbf{\widetilde{W}} \mathbf{U} = \mathbf{W} \label{eq:orb_invariance}
\end{equation}
where $\mathbf{U}$ is a unitary matrix that rotates the occupied orbitals, and $\mathbf{\widetilde{W}}$ is a feature matrix calculated in the rotated basis. As shown in detail in the Supporting Information (SI), Section S2, it is straightforward to prove that the four matrices used in this work satisfy eq~\ref{eq:orb_invariance}. This property extends to functions of these matrices, and the matrix trace is invariant to basis, 
\begin{equation}
    \Tr{\mathbf{W}_\alpha(\boldsymbol{\mathcal{\widetilde{W}}})} = \Tr{\mathbf{W}_\alpha(\boldsymbol{\mathcal{W}})}
\end{equation}
so OSMI-ACPT2 is an orbital-invariant theory.

In some cases, using matrix functions for the adiabatic connection integrand $W_\alpha(\boldsymbol{\mathcal{W}})$ may be complicated, and one could consider approximating each element $\mathbf{W}$ of $\boldsymbol{\mathcal{W}}$ by its diagonal.
This simplified approach is still size-consistent, but it is not orbital invariant---we call it orbital-based size-consistent vector interpolation (OSVI, in contrast to the matrix interpolation OSMI).
While we do not benchmark OSVI in detail in this work, it could be useful as a simple approximation to OSMI in some cases, and we have implemented it alongside OSMI in the code released with this paper (Section ~\ref{sec:comp_details}). In the SI, Section S3, we compare OSMI and OSVI to demonstrate some effects of orbital-invariance.

\subsection{Choice of Adiabatic Connection Interpolation}

A simple scalar interpolation that satisfies eqs~\ref{eq:ac_scalar_small} and~\ref{eq:ac_scalar_large} was developed by Seidl, Perdew, and Kurth~\cite{seidlSimulationAllOrderDensityFunctional2000} and named interaction-strength interpolation (ISI). However, we find that both NSC and OSMI versions of ISI produce large errors for some chemical reactions (see the SI, Section S4), so we introduce the modISI form
\begin{equation}
    W_\alpha^\text{modISI} = W_0 + \frac{\alpha W_0'}{1 - \alpha^{1/2} W_0' W_\infty' W_\text{eff}^{-2} + \alpha W_0' W_\text{eff}^{-1}} \label{eq:nsc_modisi}
\end{equation}
with $W_\text{eff}=W_\infty-W_0$, which also has a simple form and obeys the same fundamental properties of $W_\alpha$ that ISI does~\cite{seidlStrictlyCorrelatedElectrons1999,seidlSimulationAllOrderDensityFunctional2000}. Note that the denominator of eq~\ref{eq:nsc_modisi} is positive definite as long as $W_\text{eff}$ is negative and $W_\infty'$ is positive, which is always the case in this work ($W_0'$ is twice the GL2 correlation energy and therefore also always negative). Because we found that modISI is much more accurate than ISI for molecular systems (SI Section S4), the modISI adiabatic connection integrand is used for all ACPT2 calculations in the main text of this work. (Note that modISI is different than the similarly named mISI method~\cite{constantinCorrelationEnergyFunctionals2019,smigaModifiedInteractionStrengthInterpolation2020}, which modifies the models for $W_\infty$ and $W_\infty'$, rather the function $W_\alpha(\mathcal{W})$.)

Converting modISI to the OSMI formalism, we get
\begin{align}
    \mathbf{W}_\alpha^\text{modISI} &= \mathbf{W}_0 + \alpha \mathbf{W}_0' \left( \mathbf{I} + \alpha^{1/2} \mathbf{A} + \alpha \mathbf{B} \right)^{-1} \label{eq:osmi_modisi} \\
    \mathbf{A} &= -\mathbf{W}_0' \mathbf{W}_\infty' \mathbf{W}_\text{eff}^{-2} \label{eq:modisi_amat} \\
    \mathbf{B} &= \mathbf{W}_0' \mathbf{W}_\text{eff}^{-1} \label{eq:modisi_bmat}
\end{align}
If $\mathbf{W}_\text{eff}=\mathbf{W}_\infty-\mathbf{W}_0$, then eq~\ref{eq:osmi_modisi} obeys eqs~\ref{eq:ac_matrix_small} and~\ref{eq:ac_matrix_large} exactly. However, since $\mathbf{W}_\infty$ is approximated rather than exact, it is possible for this choice of $\mathbf{W}_\text{eff}$ to not be negative definite (i.e., it can have positive eigenvalues), which can cause numerical issues. Therefore, we define
\begin{align}
    \mathbf{W}_\text{eff} &= \mathbf{W}_\infty - \mathbf{W}_0 (1 - f_\text{damp}(\mathbf{W}_0\mathbf{W}_\infty^{-1})) \label{eq:mat_weff} \\
    f_\text{damp}(x) &= \frac{\ln\left(1 + \text{e}^{a(1-x)}\right)}{\ln\left(1 + \text{e}^{a}\right)}
\end{align}
which is strictly negative-definite because $\mathbf{W}_\infty$ and $\mathbf{W}_0$ are. If $a$ is large, eq~\ref{eq:mat_weff} quickly approaches $\mathbf{W}_\infty-\mathbf{W}_0$ as the eigenvalues of $\mathbf{W}_\infty$ become more negative than those of $\mathbf{W}_0$. For this work, we use $a=8$.

For the OSMI model to be numerically stable and physically realistic, the denominator of eq~\ref{eq:osmi_modisi} should be positive definite. Even though $\mathbf{W}_0'$ and 
$\mathbf{W}_\text{eff}$ are negative (semi-)definite and $\mathbf{W}_\infty'$ is positive definite, positive definiteness of the denominator is not guaranteed when using asymmetric matrices like eqs~\ref{eq:modisi_amat} and~\ref{eq:modisi_bmat} because the product of two positive definite matrices need not be positive definite. In evaluating eqs~\ref{eq:osmi_modisi}--\ref{eq:mat_weff}, we therefore use a modified definition of matrix multiplication, in which $\mathbf{AB}$ is shorthand for $\mathbf{B}^{1/2}\mathbf{A}\mathbf{B}^{1/2}$, with the latter being positive definite if $\mathbf{A}$ and $\mathbf{B}$ are.

\subsection{Strictly Correlated Electrons Limit}

Second-order gradient expansions are known for $w_\infty[n](\mathbf{r})$ and $w_\infty'[n](\mathbf{r})$ via the point change plus continuum approximation~\cite{seidlSizedependentIonizationEnergy1994,seidlStrictlyCorrelatedElectrons1999,seidlDensityFunctionalsStronginteraction2000a},
\begin{align}
    w_\infty[n](\mathbf{r}) &= An(\mathbf{r})^{4/3}(1 + \mu s^2) \label{eq:winf_gea} \\
    w_\infty'[n](\mathbf{r}) &= Cn(\mathbf{r})^{3/2}(1 + \mu' s^2) \label{eq:winfp_gea}
\end{align}
where $s=|\nabla n| / (2(3\pi^2)^{1/3} n^{4/3})$ is the reduced gradient, $A=-\frac{9}{10}\left(\frac{4\pi}{3}\right)^{1/3}$, $C=\frac{1}{2}(3\pi)^{1/2}$, $\mu=-3^{1/3}(2\pi)^{2/3}/35$, and $\mu'=-0.7222$~\cite{smigaSelfConsistentImplementationKohn2022,seidlSimulationAllOrderDensityFunctional2000,seidlDensityFunctionalsStronginteraction2000a,gori-giorgiElectronicZeroPointOscillations2009a} ($\mu'$ is the value used in \'Smiga \emph{et al.}~\cite{smigaSelfConsistentImplementationKohn2022}). In this work, to ensure that these terms do not change sign as the gradient changes (which could cause numerical issues), we use generalized gradient approximations of the form
\begin{align}
    w_\infty[n](\mathbf{r}) &= A n(\mathbf{r})^{4/3} g_\infty(s) \label{eq:modisi_winf} \\
    w_\infty'[n](\mathbf{r}) &= Cn(\mathbf{r})^{3/2} \exp(\mu' s^2) \label{eq:modisi_winfp}
\end{align}
with
\begin{equation}
    g_\infty(s) = f + (1-f) \exp\left(\frac{\mu s^2}{1 - f}\right) \label{eq:ginf}
\end{equation}
The single empirical parameter $f$ was set to $1/2$ to give reasonable dissociation curves for \ce{H2} and \ce{N2}.

\subsection{Choice of Kohn-Sham Potential}

The Kohn-Sham adiabatic connection provides the correlation energy for a given density distribution $n(\mathbf{r})$~\cite{Harris1975}, and it only provides the exact ground-state correlation energy for the exact ground-state density. In this work, as in most works using adiabatic connection GL2 models (though there are exceptions~\cite{smigaSelfConsistentImplementationKohn2022}), the correlation energy is evaluated non-self-consistently following a standard Kohn-Sham DFT calculation. As a result, the density and correlation energy are sensitive to the choice of approximate functional for the DFT calculation~\cite{constantinAdiabaticconnectionInterpolationModel2024}. In this work, for simplicity, we use the PBE functional~\cite{perdewGeneralizedGradientApproximation1996} to obtain the density, the orbitals, and the zeroth-order Hamiltonian for the GL2 energy, but this is not necessarily an optimal choice and could be improved in the future.

\subsection{Complete Specification of the ACPT2 Models}

Within either the non-size-consistent or OSMI framework for ACPT2, the model for the exchange-correlation energy is defined by the adiabatic connection integrand, the choice of initial Hamiltonian, and the strictly correlated electrons limit. Throughout this work, we use modISI, PBE, and eqs~\ref{eq:modisi_winf} and~\ref{eq:modisi_winfp} for these variables, respectively. The first two are fully nonempirical, while the latter has only one empirical parameter ($f$ in eq~\ref{eq:ginf}) and conforms to the nonempirical gradient expansions around the uniform electron gas given by eqs~\ref{eq:winf_gea} and~\ref{eq:winfp_gea}. Fully nonempirical approximations can easily be substituted if desired.~\cite{constantinCorrelationEnergyFunctionals2019,smigaModifiedInteractionStrengthInterpolation2020,smigaSelfConsistentImplementationKohn2022}

\subsection{Regularized MP2}

One of the key benefits of adiabatic connection methods is the removal of the divergence of the correlation energy as the gap vanishes. A more widely used approach to addressing this problem is a regularized MP2 theory, such as $\kappa$-MP2, whose correlation energy is~\cite{leeRegularizedOrbitalOptimizedSecondOrder2018}
\begin{equation}
    E_\text{c}^{\kappa\text{-MP2}} = \frac{1}{4} \sum_{ijab} \frac{|\mel{ij}{}{ab}|^2}{\Delta_{ij}^{ab}} \left(1 - \text{e}^{\kappa\Delta_{ij}^{ab}}\right)^2 \label{eq:kappa_mp2}
\end{equation}
with $\Delta_{ij}^{ab}=\epsilon_i+\epsilon_j-\epsilon_a-\epsilon_b$.
We introduce this method as a point of comparison for the OSMI models, since it is also size-consistent and orbital-invariant. We set $\kappa=1.1$ Ha$^{-1}$ throughout this work, which was previously found to be optimal for molecular systems~\cite{sheeRegularizedSecondOrderMoller2021a}.

\section{Results}

In Section~\ref{sec:spectator} 
we perform simple molecular calculations that demonstrate the importance of size-consistency on molecular dissociation. In Section~\ref{sec:ueg}, we show that the OSMI method provides an excellent description of the correlation energy of the uniform electron gas over a wide range of densities. Having illustrated these basic properties of OSMI, in Section~\ref{sec:res_gmtkn55} we assess its accuracy for selected subsets of the GMTKN55 database~\cite{goerigkLookDensityFunctional2017a} and compare it to common density functionals and other perturbation theory methods. We also assess the accuracy of a non-size-consistent adiabatic connection model on these databases and show that OSMI consistently matches or exceeds the accuracy of the non-size-consistent method.

\subsection{Effects of Size-Consistency on Molecular Dissociation}\label{sec:spectator}

\begin{figure}
    \centering
    \includegraphics[scale=1.0]{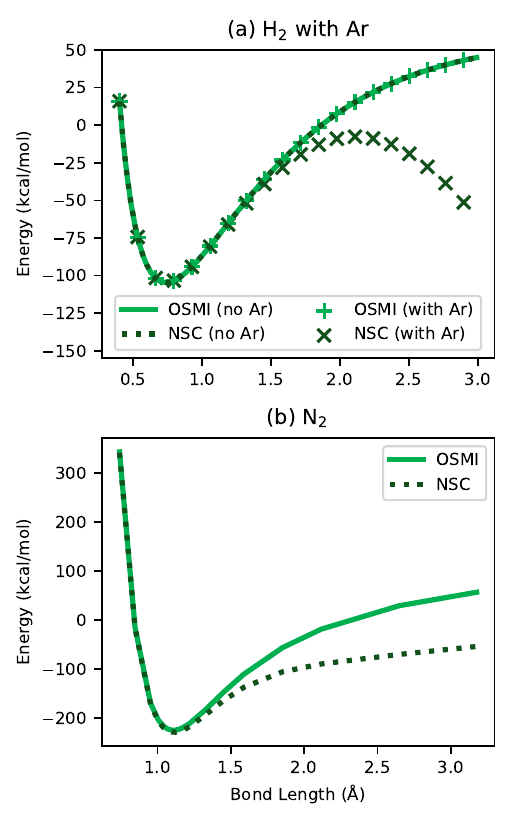}
    \caption{Spin-restricted dissociation curves of (a) the hydrogen molecule with and without an argon ``spectator atom'' 100 \r{A} away and of (b) the nitrogen molecule.  ``NSC'' is the  non-size-consistent approach of eq~\ref{eq:adiabatic_connection}, and ``OSMI'' is our size-consistent model (eq~\ref{eq:osmi_acmp2_xc}). The hydrogen dissociation curves with and without argon match for OSMI but not NSC. Energies are referenced to the spin-unrestricted dissociation limit for each method.}
    \label{fig:sc}
\end{figure}

Consider the dissociation of a molecular bond with a noble gas atom very far away. The dissociation curve should be the same as if the noble gas atom was absent, but this is only guaranteed for size-consistent models. In Figure~\ref{fig:sc}(a), we show the dissociation curve of the hydrogen molecule with and without an argon atom 100~\r{A} away. A non-size-consistent (NSC) adiabatic connection functional gives completely different dissociation curves, with terrible performance in the presence of the faraway argon atom, whereas a size-consistent functional based on the OSMI method produces identical dissociation curves in both cases, which avoid the spurious divergent behavior seen in the NSC case.

The reason for the difference is that the global elements of the $\mathcal{W}[n]$ feature vector in eq~\ref{eq:acks_nsc} are dominated by the argon atom when it is present. As a result, the adiabatic connection does not detect the static correlation regime as the hydrogen molecule dissociates, causing the energy to (nearly) diverge as the bond is dissociated in the presence of the argon atom.
This size-consistency problem can manifest even without separate ``spectator'' atoms. In Figure~\ref{fig:sc}(b)---which shows non-size-consistent and size-consistent (OSMI) models for the dissociation curve of the nitrogen molecule---the contributions to $\mathcal{W}$ from the argon electrons interfere with the correct description of the \ce{H2} electrons in the NSC case, producing an unphysical dissociation curve.

Note that these calculations are performed with frozen core electrons (see Section~\ref{sec:comp_details}). Unfreezing the core electrons further increases the contribution to $\mathcal{W}$ made by argon, resulting in even worse dissociation curves for the NSC case, as demonstrated in the SI, Section S7.

\subsection{Uniform Electron Gas}\label{sec:ueg}

\begin{figure}
    \centering
    \includegraphics[scale=1.0]{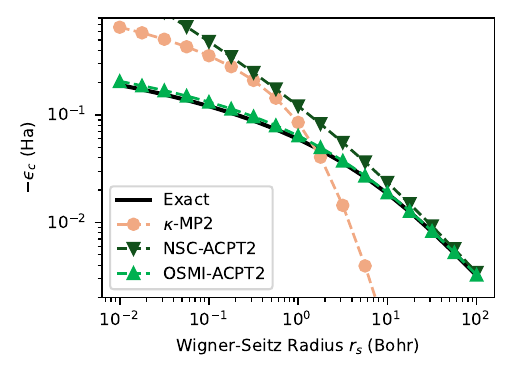}
    \caption{Correlation energy per electron $\epsilon_\text{c}$ of the uniform electron gas (UEG) for $\kappa$-MP2, non-size-consistent ACPT2 (NSC-ACPT2), and size-consistent OSMI-ACPT2. The exact result was calculated using the modified PW92 LDA functional~\cite{perdewAccurateSimpleAnalytic1992} (LDA\_C\_PW\_MOD in libxc~\cite{lehtolaRecentDevelopmentsLibxc2018}), which interpolates over quantum Monte Carlo results for the correlation energy of the UEG.~\cite{ceperleyGroundStateElectron1980}}
    \label{fig:ueg}
\end{figure}

The uniform electron gas (UEG), a simple model of bulk metals, is completely defined by its Wigner-Seitz radius $r_\text{s}=(4\pi n/3)^{-1/3}$, where $n$ is the density. In three dimensions, second-order perturbation theories diverge due to a conspiracy of the gapless spectrum, the finite density of states at the Fermi energy, and the long-range nature of the Coulomb interaction (see the SI, Section S1, for further discussion).
The divergence of the GL2 correlation energy means that non-size-consistent models for $W_\alpha$ of the ISI or modISI form (eq 9 of Seidl \emph{et al.}~\cite{seidlSimulationAllOrderDensityFunctional2000} and eq~\ref{eq:nsc_modisi} of this work, respectively) have the limit
\begin{equation}
    \lim_{W_0'\rightarrow\infty} W_\alpha^\text{(mod)ISI} = \frac{\alpha^{1/2}W_\text{eff}}{\alpha^{1/2} - W_\infty'/W_\text{eff}} \,\,\,\,\,\,\,\,\,\,\,  (\alpha > 0) \label{eq:modisi_w0p_to_inf}
\end{equation}
which holds for all $\alpha>0$. As $r_\text{s} \rightarrow 0$, these models predict that the correlation energy per electron goes as $\epsilon_\text{c} \sim r_\text{s}^{-1/2}$ (using eqs~\ref{eq:acks_nsc},~\ref{eq:winf_gea}, and~\ref{eq:winfp_gea}), which does not match the known high-density limit $\epsilon_\text{c} \sim \ln{r_\text{s}}$. Therefore, as shown in Figure~\ref{fig:ueg}, non-size-consistent ISI/modISI accurately describes the low-density limit but drastically overestimates the magnitude of the correlation energy in the high-density limit.

In contrast, the size-consistent OSMI model is accurate over the entire range of $r_s$, quantitatively recovering the logarithmic scaling at small $r_s$. This improved performance is because the OSMI method performs the adiabatic connection separately for each electron. For the UEG, the second-order correlation energy only diverges for electrons near the Fermi level, so electrons far from the Fermi level can reasonably be treated with a weak correlation model, like GL2.
In Figure~\ref{fig:ueg}, we also compare the behavior of OSMI-ACPT2 to $\kappa$-MP2 and find that $\kappa$-MP2 only provides a reasonable correlation energy at a small range of intermediate densities, $r_s \sim 1\text{--}3$. (Interestingly, this range is not far from typical valence electron densities of simple metals.)

The UEG also illustrates a problem with the real-space local interpolation approach~\cite{zhouConstructionExchangecorrelationFunctionals2015,vuckovicExchangeCorrelationFunctionals2016,kooiLocalGlobalInterpolations2018,daasExactMollerPlessetAdiabatic2025} for size-consistent adiabatic connection functionals. By symmetry, the $w(\mathbf{r})$ energy densities are the same for all $\mathbf{r}$, so local interpolation will yield the same unphysical result as NSC-ACPT2 in Figure~\ref{fig:ueg}. By treating single-particle states separately for the interpolation, as opposed to treating real-space coordinates separately, OSMI avoids this problem.

\subsection{Molecular Benchmarks}\label{sec:res_gmtkn55}

\begin{figure*}
    \centering
    \includegraphics[scale=1.0]{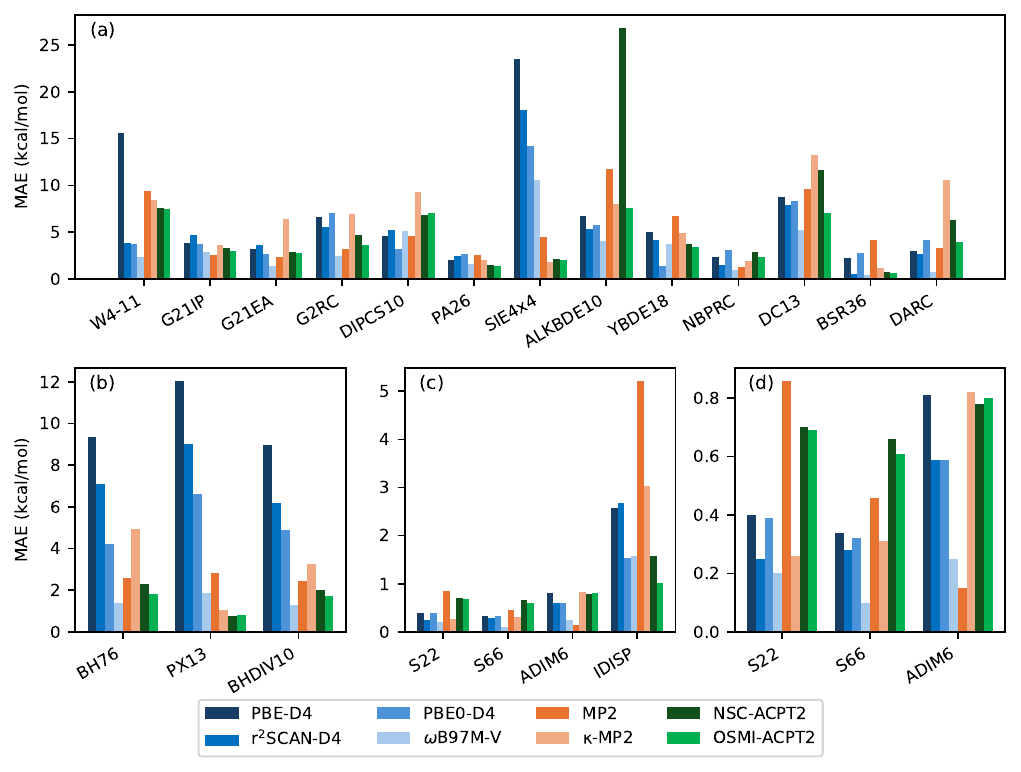}
    \caption{Mean absolute errors (MAEs) of density functionals and MP2/GL2 variants for selected subsets of the GMTKN55 database. The subsets are separated based on type of property: (a) reaction energies, ionization potentials, and electron affinities, (b) barrier heights, (c) noncovalent interactions, and (d) noncovalent interactions excluding IDISP, to make the energy scale more readable.}
    \label{fig:gmtkn55}
\end{figure*}

\begin{table}[]
    \caption{Average errors of different methods over the 20 sub-databases in Figure~\ref{fig:gmtkn55}, in kcal mol$^{-1}$.}
    \centering
    \begin{tabular}{crrr}
        \hline
        Method & MoM Error$^a$ & WTMAD-2 Error$^b$ \\
        \hline
        PBE-D4 & 6.10 & 8.75 \\
        r$^2$SCAN-D4 & 4.57 & 6.24 \\
        PBE0-D4 & 4.05 & 5.36 \\
        $\omega$B97M-V & 2.40 & 2.28 \\
        MP2 & 4.02 & 4.93 \\
        $\kappa$-MP2 & 4.60 & 5.48 \\
        NSC-modISI & 4.49 & 4.35 \\
        OSMI-modISI & 2.97 & 3.53 \\
        \hline
    \end{tabular}\\
    $^a$Mean-of-means error, i.e.\ the mean of the mean absolute error over all 20 databases.
    $^b$The weighted average error metric defined by eq~\ref{eq:wtmad2} and introduced by Goerigk \emph{et al.}~\cite{goerigkLookDensityFunctional2017a}
    \label{tab:avg_errs}
\end{table}

\begin{table}[]
    \caption{Mean absolute errors on the W4-11 atomization energy dataset for different methods, with and without core correlation, in kcal mol$^{-1}$.}
    \centering
    \begin{tabular}{crrr}
        \hline
        Method & Frozen core$^a$ & Correlated core$^a$ & Difference \\
        \hline
        PBE-D4 & 15.63 & 15.98 & 0.34 \\
        r$^2$SCAN-D4 & 3.78 & 4.05 & 0.26 \\
        PBE0-D4 & 3.68 & 3.51 & -0.17 \\
        $\omega$B97M-V & 2.33 & 2.19 & -0.14 \\
        MP2 & 9.38 & 10.33 & 0.95 \\
        $\kappa$-MP2 & 8.44 & 8.00 & -0.44 \\
        NSC-ACPT2 & 7.55 & 12.81 & 5.26 \\
        OSMI-ACPT2 & 7.45 & 6.74 & -0.70 \\
        \hline
    \end{tabular}\\
    $^a$``Frozen core'' refers to calculations with the aug-cc-pVQZ basis with density fitting, while ``Correlated core'' refers to calculations with the aug-cc-pCVQZ basis without density fitting. Both methods use $x^{-3}$ basis set extrapolation with X=3,4 for the correlation term of wave function methods, as described in Section~\ref{sec:comp_details}.
    \label{tab:core_corr}
\end{table}

Having explored some of the key properties of the OSMI model, we now benchmark its performance for main-group chemistry and compare to other methods. Figure~\ref{fig:gmtkn55} shows the mean absolute errors (MAEs) of several density functional approximations and second-order perturbation theory methods for selected subsets of the GMTKN55 database~\cite{goerigkLookDensityFunctional2017a}, and Table~\ref{tab:avg_errs} shows the average errors over these subsets. The WTMAD-2 is a weighted error metric designed to give similar weight to reaction energies with large and small energy magnitudes and is given by~\cite{goerigkLookDensityFunctional2017a}
\begin{equation}
    \text{WTMAD-2} = \frac{1}{\sum_{i=1}^{20} N_i} \sum_{i=1}^{20} N_i \frac{56.84 \text{ kcal mol}^{-1}}{\widebar{|\Delta E|}_i} \cdot \text{MAE}_i \label{eq:wtmad2}
\end{equation}
where $i$ indexes the 20 subsets of GMTKN55 computed in this work, $N_i$ is the number of data points in subset $i$, $\widebar{|\Delta E|}_i$ is the mean absolute value of the reference values in subset $i$, and $\text{MAE}_i$ is the mean absolute error of subset $i$ compared to the high-level quantum chemistry reference values.

Our first key observation is that OSMI-ACPT2 is always at least comparable to the conventional, non-size-consistent evaluation of the same adiabatic connection (NSC-ACPT2), and in a few cases, such as group-1 and group-2 bond dissociation energies (ALKBDE10), OSMI-ACPT2 is significantly more accurate than NSC-ACPT2. This shows that size-consistency is critical not just for the model systems in Section~\ref{sec:spectator}, but also for general chemistry problems. In addition, Table~\ref{tab:avg_errs} shows that OSMI-ACPT2 is on average more accurate than both canonical MP2 and $\kappa$-MP2 for these benchmarks, and it also outperforms the nonempirical density functionals PBE~\cite{perdewGeneralizedGradientApproximation1996}, r$^2$SCAN~\cite{furnessAccurateNumericallyEfficient2020}, and PBE0~\cite{adamoReliableDensityFunctional1999,ernzerhofAssessmentPerdewBurke1999}, all three of which were evaluated with semi-empirical D4 dispersion corrections~\cite{caldeweyherExtensionD3Dispersion2017,caldeweyherGenerallyApplicableAtomiccharge2019a} to improve their accuracy for non-covalent interactions. The only method tested that outperforms OSMI-ACPT2 on average is the empirically parametrized $\omega$B97M-V functional.~\cite{mardirossianOB97MVCombinatoriallyOptimized2016}

Beyond its good average performance across these benchmarks, OSMI-ACPT2 provides a much more accurate description of the SIE4x4 dataset---which consists of reactions prone to DFT self-interaction error---than any of the density functionals investigated here, while also predicting barrier heights with an accuracy on par with $\omega$B97M-V,~\cite{mardirossianOB97MVCombinatoriallyOptimized2016} (which is among the most accurate available functionals for these benchmarks~\cite{mardirossianThirtyYearsDensity2017}). Because barrier heights require the description of transition states and therefore stretched bonds with some measure of static correlation, this result indicates that OSMI-ACPT2 can simultaneously mitigate self-interaction error and static correlation error. In contrast, for DFT, there is a well-established trade-off between these two types of error~\cite{cohenChallengesDensityFunctional2012a} (typically only overcome by sophisticated local hybrid functionals~\cite{kauppNextGenerationDensity2024b}).

Our benchmarks also reveal a key drawback of NSC-ACPT2 compared to size-consistent methods. The results of Figure~\ref{fig:gmtkn55} were obtained with frozen core electrons for PT2 methods, as is common practice in quantum chemistry because core electrons are not expected to contribute significantly to chemical bonding. Indeed, for the W4-11 atomization energies, Table~\ref{tab:core_corr} shows that all methods tested here have similar mean absolute errors (MAE) regardless of whether core electrons are frozen, with the critical exception of NSC-ACPT2. While the other methods see MAE differences of less than 1 kcal mol$^{-1}$, NSC-ACPT2 is 5 kcal mol$^{-1}$ \emph{less} accurate when core correlation effects are included. This is a symptom of non-size-consistency, and it suggests that the accuracy of NSC-ACPT2 will be reliant on excluding core correlation, especially for larger atoms. This is an unphysical and undesirable property of a model chemistry that is completely avoided by OSMI-ACPT2.

\begin{figure*}
    \centering
    \includegraphics[scale=1.0]{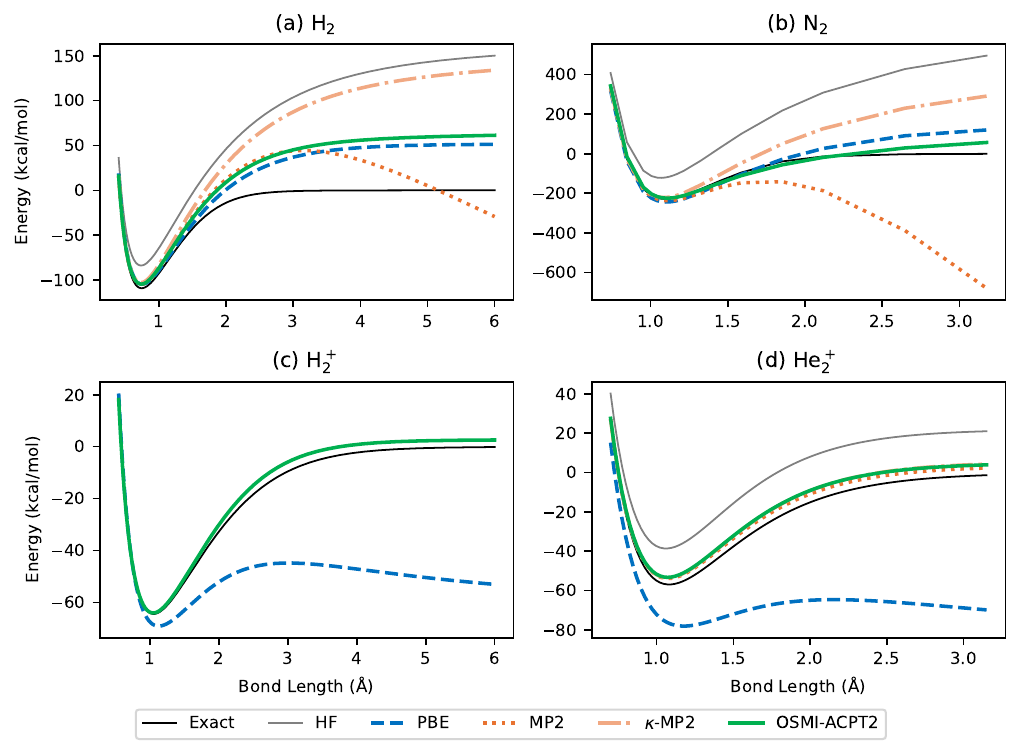}
    \caption{Spin-restricted dissociation curves of (a) the hydrogen molecule and (b) the nitrogen molecule with different methods, as well as doublet dissociation curves of (c) the \ce{H2+} molecular ion and (d) the \ce{He2+} molecular ion. Energies are referenced to the open-shell dissociation limit for each method. The ``exact'' result for \ce{H2} was obtained from CCSD in the aug-cc-pVQZ basis~\cite{dunningGaussianBasisSets1989,kendallElectronAffinitiesFirstrow1992,woonGaussianBasisSets1993}, while for nitrogen it was taken from Gdanitz~\cite{gdanitzAccuratelySolvingElectronic1998}, which used multi-reference coupled-cluster theory. The ``exact'' result for \ce{H2+} was obtained from Hartree-Fock, while for \ce{He2+} it was obtained from CCSD(T), with both methods using the aug-cc-pVQZ basis. MP2 and $\kappa$-MP2 are not shown for \ce{H2+} because they are exact for 1-electron systems. For \ce{He2+}, MP2, $\kappa$-MP2, and OSMI-ACPT2 produce very similar dissociation curves.}
    \label{fig:diss}
\end{figure*}

Before concluding, we revisit the dissociation curves of the hydrogen and nitrogen molecules in addition to the dissociation of the \ce{H2+} and \ce{He2+} ions. In Figures~\ref{fig:diss}(a) and~\ref{fig:diss}(b), we see the well-known divergence of MP2 as the hydrogen-hydrogen and nitrogen-nitrogen bonds are stretched and the highest occupied and lowest unoccupied orbitals become degenerate. $\kappa$-MP2 does not suffer from this problem, but it drastically overestimates the energy at large bond lengths because it cannot describe the static correlation of the stretched bond. In fact, $\kappa$-MP2 is not much better than Hartree-Fock for the dissociation of the hydrogen molecule. OSMI-ACPT2 also does not fully capture the static correlation at large bond lengths, but it is much closer to the exact result than Hartree-Fock and $\kappa$-MP2, and its description of the stretched bonds is on par with PBE. As show in Figures~\ref{fig:diss}(c) and~\ref{fig:diss}(d), this improvement comes without the large self-interaction error suffered by PBE for problems like stretched \ce{H2+}~\cite{cohenChallengesDensityFunctional2012a} and \ce{He2+}. Because OSMI-ACPT2 (eq~\ref{eq:osmi_modisi}) predicts no correlation for one-electron systems, it would describe \ce{H2+} dissociation exactly if the exact Kohn-Sham potential was used for the mean-field Hamiltonian. In practice, an approximate mean-field Hamiltonian must be used (PBE in this work), resulting in some density-driven error. However, as shown in Figure~\ref{fig:diss}(c), this error is quite small for \ce{H2+} compared to PBE, which exhibits pathological self-interaction error for this problem. A similar trend is observed for \ce{He2+} (Figure~\ref{fig:diss}(d)). Note that MP2 and $\kappa$-MP2 are not plotted for \ce{H2+} because they are exact for this system. In addition, MP2 and $\kappa$-MP2 are very similar to OSMI-ACPT2 for \ce{He2+}, so we provide a clearer plot of the difference between these methods in the SI, Section S8.

\section{Conclusion}

We have introduced a size-consistent variant of adiabatic connection functionals that is orbital-invariant, does not require the identification of molecular fragments, and does not require the explicit evaluation of the PT2 correlation energy density. The resulting OSMI method cures some key shortcomings of adiabatic connection functionals, paving the way for broader applications of this promising class of model chemistries. To demonstrate the method, we introduced an adiabatic connection model called modISI. Paired with the OSMI method, modISI predicts more accurate thermochemical properties than MP2 and $\kappa$-MP2, and is on par with state-of-the-art empirical hybrid density functionals for barrier heights.

Some challenges remain in the design of these adiabatic connection models. For example, atomization energies with OSMI-ACPT2 are not particularly impressive compared to meta-GGA and hybrid density functionals, and this might be improved by more careful design of the adiabatic connection formula and/or the strong interaction limit functionals. In addition, since our method uses the Kohn-Sham adiabatic connection, it assumes access to the exact Kohn-Sham potential, which is unknown. Our use of the PBE potential in place of the exact one is certainly a coarse approximation that could lead to significant density-driven error in some cases. This problem could be addressed by better density functionals, use of the optimized effective potential method~\cite{constantinAdiabaticconnectionInterpolationModel2024}, or a self-consistent approach to evaluating the adiabatic connection functionals~\cite{smigaSelfConsistentImplementationKohn2022}. Finally, we note that some additional methodological choices, such as including single-excitation terms in the PT2 energy~\cite{gorlingCorrelationenergyFunctionalIts1993,fabianoInteractionStrengthInterpolationMethod2016} and using a restricted-orbital formalism for open-shell systems,~\cite{knowlesRestrictedMollerPlesset1991} have been explored for other PT2 methods and could be tested for their potential to improve OSMI-ACPT2 as well.

Looking forward, the development of an intrinsically size-consistent adiabatic connection functional formalism opens the door to studying large systems---including molecules, solids, and interfaces---since the physics of different subsystems can be described within one framework. In particular, the simultaneous accuracy of OSMI-ACPT2 for barrier heights and the uniform electron gas encourages applications to surface chemistry and heterogeneous catalysis. Despite the apparently high computational cost of our method, which scales as $O(N^5)$ due to the GL2 correlation energy, its orbital invariance guarantees the success of local correlation approximations that can significantly lower the cost.

\section{Computational Details} \label{sec:comp_details}

All molecular calculations were performed using the PySCF software package~\cite{sunPySCFPythonbasedSimulations2018,sunRecentDevelopmentsPySCF2020,sunLibcintEfficientGeneral2015}. Unless otherwise specified, we used the aug-cc-pVQZ basis~\cite{dunningGaussianBasisSets1989,kendallElectronAffinitiesFirstrow1992,woonGaussianBasisSets1993} with density fitting for all molecular calculations. We used the ``rifit'' density fitting basis~\cite{weigendEfficientUseCorrelation2002} for the Coulomb, exchange, and correlation terms. For comparison, we also computed the GMTKN55 benchmarks with the def2-QZVPPD basis~\cite{weigendGaussianBasisSets2003,rappoportPropertyoptimizedGaussianBasis2010} and used density fitting with the def2-QZVPPD-RI basis.~\cite{hattigOptimizationAuxiliaryBasis2005,hellwegDevelopmentNewAuxiliary2014} The def2-QZVPPD basis results can be found in the SI, Section S6, and are overall quite similar to those with the aug-cc-pVQZ results. For the bond dissociation curve calculations of Figures~\ref{fig:sc} and~\ref{fig:diss}, the aug-cc-pVQZ basis was used without density fitting. We found the basis set exchange~\cite{fellerRoleDatabasesSupport1996,pritchardNewBasisSet2019,schuchardtBasisSetExchange2007} useful for selecting basis sets.

For the GMTKN55 benchmarks (but not the dissociation curves), we used the $x^{-3}$ extrapolation technique~\cite{helgakerBasissetConvergenceCorrelated1997} with $x=3,4$ to extrapolate the MP2, $\kappa$-MP2, NSC-ACPT2, and OSMI-ACPT2 correlation energies to the complete basis set limit from aug-cc-pVTZ and aug-cc-pVQZ calculations (or def2-TZVPPD and def2-QZVPPD calculations, in the case of Section S6 of the SI). All other quantities were taken from the quadruple-zeta basis results and not extrapolated. For the S22, S66, and ADIM6 datasets, counterpoise corrections were applied to mitigate basis set superposition error. For the elements Ca and K, the aug-cc-pVQZ basis was not available, so we used def2-QZVPPD for these elements. The ALKBDE10, G21IP, and G2RC datasets were evaluated without density fitting for aug-cc-pVQZ because the corresponding ``rifit'' basis was not available for the elements Li and Na. All MP2 and ACPT2 calculations were performed with frozen core electrons unless otherwise specified. For calculations of the W4-11 atomization energies and for the bond dissociation curves in Section S7 of the SI, the aug-cc-pCVTZ and aug-cc-pCVQZ basis sets were used to introduce basis functions for core correlation,~\cite{woonGaussianBasisSets1995,petersonAccurateCorrelationConsistent2002} and density fitting was not used.

It has been observed that for open-shell systems, the spin-contamination of the Hartree-Fock wave function can degrade the accuracy of MP2.~\cite{sheeRegularizedSecondOrderMoller2021a} For this reason, we used a restricted-orbital formalism called RMP2~\cite{knowlesRestrictedMollerPlesset1991} for all MP2 and $\kappa$-MP2 results. Unlike for the doubles contribution (eq~\ref{eq:kappa_mp2}), we did not use any regularization on the resulting singles contribution for $\kappa$-MP2. In the SI, Section S5, we compare unrestricted MP2 to RMP2 and find that RMP2 is much more accurate for the BH76 dataset and otherwise comparable to unrestricted MP2. Because the orbitals of a DFT calculation are fictional, rather than an approximate ansatz for the wave function like in Hartree-Fock, spin contamination is a less important concept for GL2 than for MP2, so all ACPT2 calculations were performed within the spin-unrestricted formalism. However, one could in principle also use a restricted-orbital formalism for ACPT2.

The default (level 3) integration grids in PySCF were used to integrate density functionals (with the exception of the \ce{H2+} PBE calculations, which used level 5 grids), including the matrix elements eq~\ref{eq:wdf_eqn} for the OSMI-ACPT2 method, and libxc~\cite{lehtolaRecentDevelopmentsLibxc2018} was used to evaluate standard semilocal functionals. For ACPT2, the adiabatic connections of eq~\ref{eq:adiabatic_connection} and eq~\ref{eq:osmi_acmp2_xc} were integrated numerically over $\alpha$ using the midpoint rule with a uniform grid of 512 points.

We have released a PySCF extension along with this paper that can be used to run the ACPT2 calculations, and it is available at \url{https://github.com/kylebystrom/mp2-variants}. Note that in the code, the method is called ``ACMP2'' rather than ``ACPT2'', even though it can be used for both GL2 and MP2-based methods. Also, in the package, the RMP2 method is called ``ROMP2'' to avoid confusion with spin-restricted, closed-shell MP2, which is called RMP2 in PySCF.

The uniform electron gas calculations where performed via numerical integration of the MP2 equations in the Brillouin Zone. The code for performing these calculations is also available at \url{https://github.com/kylebystrom/mp2-variants}. See the SI, Section S1, for the transformation of the MP2 equations into integrals in reciprocal space and for the grids used for numerical quadrature. Section S1 also provides an analysis of the divergence of GL2 for the uniform electron gas and a basic benchmark of the precision of the numerical quadrature.

The scripts for performing the calculations and generating the figures in this work are provided in a repository at \url{https://github.com/kylebystrom/osmi_2025_scripts}. Running these scripts requires access to the GMTKN55 data~\cite{goerigkLookDensityFunctional2017a} in a yaml file format, which we provide at the following additional repository: \url{https://github.com/kylebystrom/YQCC}, which was built from the repository for the ACCDB collection of chemistry databases~\cite{morganteACCDBCollectionChemistry2019}. The data computed for this paper will be published to Zenodo upon publication of the final manuscript.

\section*{Associated Content}
\subsection*{Supporting Information}

The Supporting Information is available free of charge at [link].
Theory and methods for uniform electron gas calculations; proof that OSMI-ACPT2 is orbital-invariant; comparisons between OSVI and OSMI, ISI and modISI, MP2 and RMP2, and different basis sets; and additional dissociation curve plots.

\section*{Author Information}

\subsection*{Notes}
The authors declare no competing financial interest.

\section*{Acknowledgments}
The Flatiron Institute is a division of the Simons Foundation.

\bibliography{mybib}

\end{singlespace}
\end{document}


\preprint{APS/123-QED}

\date{\today}

\onecolumngrid

\setcounter{page}{1}

\MakeTitle{Supplementary Information for ``Size-Consistent Adiabatic Connection Functionals via Orbital-Based Matrix Interpolation''}{Kyle Bystrom, Timothy Berkelbach}

\beginsupplement

\section{PT2 in the Uniform Electron Gas at the Thermodynamic Limit}

In the uniform electron gas, the (spatial symmetry-restricted) mean-field states are plane-waves
\begin{equation}
    \phi_\mathbf{k}(\mathbf{r}) = \frac{1}{\sqrt{\Omega}} \text{e}^{i\mathbf{k}\cdot\mathbf{r}}
\end{equation}
where $\Omega$ is the volume of the periodic cell, and we take $\Omega\rightarrow\infty$ to reach the thermodynamic limit. The zero-temperature, mean-field ground-state $\Phi_0$ consists of filling all states with $|\mathbf{k}| < k_\text{F}$, with
\begin{equation}
    \frac{1}{n} = \frac{4\pi}{3} r_\text{s}^3 = 3\pi^2\frac{1}{k_\text{F}^3} \label{eq:rho_rs_kf}
\end{equation}
defining the density $n$, Wigner-Seitz radius $r_\text{s}$, and Fermi wave-vector $k_\text{F}$. The direct Coulomb matrix element between the ground state and this double excitation is simply
\begin{equation}
    \mel{\Phi_0}{}{\Phi_{\mathbf{k},\mathbf{p}}^{\mathbf{k}+\mathbf{q}, \mathbf{p} - \mathbf{q}}}_\text{d} = v(q) = \frac{1}{\Omega} \frac{4\pi}{q^2}
\end{equation}
while the exchange part is
\begin{equation}
    \mel{\Phi_0}{}{\Phi_{\mathbf{k},\mathbf{p}}^{\mathbf{k}+\mathbf{q}, \mathbf{p} - \mathbf{q}}}_\text{x} = \delta_{\sigma_\mathbf{k}\sigma_\mathbf{p}} v(|\mathbf{k+p-q}|) = -\frac{\delta_{\sigma_\mathbf{k}\sigma_\mathbf{p}}}{\Omega} \frac{4\pi}{|\mathbf{k+q-p}|^2}
\end{equation}
where atomic units are used throughout, and where the wave-vectors $\mathbf{k},\mathbf{p}$ implicitly contain spins $\sigma_{\mathbf{k}},\sigma_{\mathbf{p}}$. All the matrices of $\boldsymbol{\mathcal{W}}$ are diagonal for the uniform electron gas (due to the uniformity of the density for density functionals, and the conservation of momentum selection rule for PT2), so we only need to find the diagonals, which are given by
\begin{align}
    \left(\mathbf{W}_0'\right)_{\mathbf{kk'}} = \delta_{\mathbf{k}\mathbf{k'}} W_0'(k) &= -\frac{1}{2} \sum_\mathbf{p} \sum_\mathbf{q} \frac{\Theta(k_\text{F}-k) \Theta(k_\text{F}-p) \Theta(|\mathbf{k+q}|-k_\text{F}) \Theta(|\mathbf{p-q}|-k_\text{F})}{\epsilon_\mathbf{|k+q|} + \epsilon_\mathbf{|p-q|} - \epsilon_k - \epsilon_p} (v(q) - \delta_{\sigma_\mathbf{k}\sigma_\mathbf{p}} v(|\mathbf{k+q-p}|))^2
\end{align}
The Heaviside step function terms enforce that two-particle excitations occur only from occupied states and only to unoccupied states. Also note that due to symmetry, $W_0'(k)$ is independent of the direction of $\mathbf{k}$, so we assume that it points in the $\hat{z}$ direction.

In this work, we examine only the non-spin-polarized electron gas. Reducing the above equation with spin symmetry gives
\begin{equation}
    W_0'(k) = -\sum_\mathbf{p} \sum_\mathbf{q} \frac{\Theta(k_\text{F}-k) \Theta(k_\text{F}-p) \Theta(|\mathbf{k+q}|-k_\text{F}) \Theta(|\mathbf{p-q}|-k_\text{F})}{\epsilon_\mathbf{|k+q|} + \epsilon_\mathbf{|p-q|} - \epsilon_k - \epsilon_p} (2 v(q)^2 - v(q)v(|\mathbf{k+q-p}|))
\end{equation}
where now the wave-vectors $\mathbf{k},\mathbf{p}$ do not contain a spin index and merely denote the spatial plane-wave orbitals. To evaluate this equation in the thermodynamic limit, we convert the sums over plane-waves to integrals
\begin{align}
    W_0'(k) &= -\left(\frac{\Omega}{8\pi^3}\right)^2 \int \dd[3]\mathbf{p} \int \dd[3]\mathbf{q} \, \frac{\Theta(k_\text{F}-k) \Theta(k_\text{F}-p) \Theta(|\mathbf{k+q}|-k_\text{F}) \Theta(|\mathbf{p-q}|-k_\text{F})}{\epsilon_\mathbf{|k+q|} + \epsilon_\mathbf{|p-q|} - \epsilon_k - \epsilon_p} (2 v(q)^2 - v(q)v(|\mathbf{k+q-p}|))\\
    &=-\left(\frac{1}{2\pi^2}\right)^2\int \dd[3]\mathbf{p} \int \dd[3]\mathbf{q} \, \frac{\Theta(k_\text{F}-k) \Theta(k_\text{F}-p) \Theta(|\mathbf{k+q}|-k_\text{F}) \Theta(|\mathbf{p-q}|-k_\text{F})}{\epsilon_\mathbf{|k+q|} + \epsilon_\mathbf{|p-q|} - \epsilon_k - \epsilon_p} (2 V(q)^2 - V(q) V(|\mathbf{k+q-p}|))
\end{align}
with $V(q)=1/q^2$.

We now transform the intregrals into spherical coordinates. This gives an integral over the norms of $\mathbf{q}$ and $\mathbf{p}$, their polar angles $\theta_\mathbf{q}$ and $\theta_\mathbf{p}$, and their azimuthal angles $\phi_\mathbf{q}$ and $\phi_\mathbf{p}$. Except for the exchange Coulomb interaction $V(|\mathbf{k+q-p}|)$, the integral is independent of the angle between $\mathbf{k}=k\hat{z}$ and $\mathbf{p}$. Therefore, we do a change of variables to integrate over
\begin{itemize}
    \item $\theta_{kq}$: The polar angle between $\mathbf{k}$ and $\mathbf{q}$.
    \item $\theta_{qp}$: The polar angle between $\mathbf{q}$ and $\mathbf{p}$.
    \item $\phi_{kq}$: The azimuthal angle of $\mathbf{q}$ around the axis of $\mathbf{k}$.
    \item $\phi_{kqp}$: The azimuthal angle of $\mathbf{p}$ around the axis of $\mathbf{q}$, minus the azimuthal angle of $\mathbf{k}$ around the axis of $\mathbf{q}$.
\end{itemize}
This change of variables yields
\begin{align}
    W_0'(k) = -\left(\frac{1}{2\pi^2}\right)^2 \int_0^{k_\text{F}} & p^2\dd p \int_0^\infty q^2\dd q \, \int_{-1}^1 \dd\left(\cos\theta_{kq}\right) \int_{-1}^1 \dd\left(\cos\theta_{qp}\right) \int_0^{2\pi} \dd\phi_{kq} \int_0^{2\pi} \dd\phi_{kqp} \notag \\
    &\times \frac{\Theta(k^2+q^2+2kq\cos(\theta_{kq})-k_\text{F}^2) \Theta(p^2+q^2-2pq\cos(\theta_{qp})-k_\text{F}^2)}{\epsilon_{\sqrt{k^2+q^2+2kq\cos(\theta_{kq})}} + \epsilon_{\sqrt{p^2+q^2-2pq\cos(\theta_{qp})}} - \epsilon_k - \epsilon_p} \notag\\
    &\times (2 V(q)^2 - V(q) V(Q))
\end{align}
Note that the $\Theta(k_\text{F}-p)$ has been absorbed into the integration bound on $p$, and the $\Theta(k_\text{F}-k)$ is implied because $W_0'(k)$ is computed only for occupied states. Also, we have introduced the short-hand
\begin{align}
    Q &= |\mathbf{k-p+q}| = \sqrt{Q_1 - Q_2\cos\phi_{kqp}} \\
    Q_1 &= (k\cos\theta_{kq}-p\cos\theta_{qp}+q)^2 + k_\perp^2 + p_\perp^2 \\
    Q_2 &= 2k_\perp p_\perp \cos\phi_{kqp} \\
    k_\perp &= k\sin\theta_{kq} \\
    p_\perp &= p\sin\theta_{qp}
\end{align}
Next, for computational simplicity, we scale the radial coordinates by $k_\text{F}$. Also, by symmetry, integrating over $\phi_{kq}$ must give a trivial factor of $2\pi$:
\begin{align}
    W_0'(k_\text{F}k) = -\frac{1}{2\pi^3} \int_0^1 & p^2\dd p \int_0^\infty q^2\dd q \, \int_{-1}^1 \dd\left(\cos\theta_{kq}\right) \int_{-1}^1 \dd\left(\cos\theta_{qp}\right) \int_0^{2\pi} \dd\phi_{kqp} \notag \\
    &\times \frac{\Theta(k^2+q^2+2kq\cos(\theta_{kq})-1) \Theta(p^2+q^2-2pq\cos(\theta_{qp})-1)}{E(\sqrt{k^2+q^2+2kq\cos(\theta_{kq})}) + E(\sqrt{p^2+q^2-2pq\cos(\theta_{qp})}) - E(k) - E(p)} \notag\\
    &\times (2 V(q)^2 - V(q)V(Q))
\end{align}
where we have introduced the reduced eigenvalues
\begin{equation}
    E(k) = \frac{\epsilon_{k_\text{F}k}}{{k_\text{F}^{}}^2}
\end{equation}
Also note that the input wave-vector on the left-hand side is now scaled by $k_\text{F}$.

Next, we must integrate $\phi_{kqp}$. Using Mathematica~\cite{Mathematica}, we obtain the integral relation
\begin{equation}
    \int_0^{2\pi} d\phi \frac{1}{Q_1 - Q_2\cos\phi} = \frac{2\pi}{\sqrt{Q_1^2-Q_2^2}} = \frac{2\pi}{\tilde{Q}^2}
\end{equation}
where $\tilde{Q} = (Q_1^2-Q_2^2)^{1/4}$. Using this relation to integrate $\phi_{kqp}$, and also replacing the remaining Heaviside step functions with bounds on the polar angle integrals, we obtain
\begin{align}
    W_0'(k_\text{F}k) = -\frac{1}{\pi^2} \int_0^1 & p^2\dd p \int_0^\infty q^2\dd q \, \int_{x_-(k,q)}^{1} \dd x \int_{-1}^{x_+(p,q)} \dd y \notag \\
    &\times \frac{1}{E(\sqrt{k^2+q^2+2kqx}) + E(\sqrt{p^2+q^2-2pqy}) - E(k) - E(p)} \notag\\
    &\times \left(\frac{2}{q^4} - \frac{1}{q^2\tilde{Q}^2}\right)
\end{align}
with the bounds given by
\begin{align}
    x_-(k,q) &= \max\left(\frac{1 - k^2 - q^2}{2kq}, -1\right) \\
    x_+(k,q) &= \min\left(\frac{-1 + k^2 + q^2}{2kq}, 1\right)
\end{align}
Finally, we separate the radial and angular integrals
\begin{align}
    W_0'(k) &= -\frac{1}{\pi^2} \int_0^1 p^2\dd p \int_0^\infty q^2\dd q \, I(k/k_\text{F},p,q) \label{eq:w0p_ueg} \\
    I(k,p,q) &= \int_{x_-(k,q)}^{1} \dd x \int_{-1}^{x_+(p,q)} \dd y \frac{2q^{-4} - q^{-2}\tilde{Q}^{-2}}{\Delta(k,p,q,x,y)} \label{eq:ikpq} \\
    \Delta(k,p,q,x,y) &= E(\sqrt{k^2+q^2+2kqx}) + E(\sqrt{p^2+q^2-2pqy}) - E(k) - E(p)
\end{align}
This gives us a general formula for the elements of the diagonal PT2 energy matrix $\mathbf{W}_0'$ that can be applied to various choices of mean-field initial Hamiltonian simply by modifying the definition of the eigenvalues.

For $\kappa$-MP2, we use the Hartree-Fock ground-state Hamiltonian and use
\begin{equation}
    I^{\kappa\text{-MP2}}(k,p,q) = \int_{x_-(k,q)}^{1} \dd x \int_{-1}^{x_+(p,q)} \dd y \frac{2q^{-4} - q^{-2}\tilde{Q}^{-2}}{\Delta(k,p,q,x,y)} \left[1 - \exp(-\kappa \Delta(k,p,q,x,y))\right]^2
\end{equation}
in place of $I(k,p,q)$ when evaluating eq~\ref{eq:w0p_ueg}.

\subsection{Evaluation of ACPT2}

The ACPT2 correlation energy is
\begin{align}
    E_\text{c}^\text{ACPT2} &= \int_0^1 \dd\alpha W_\alpha(\boldsymbol{\mathcal{W}})
\end{align}
where the integral over $\mathbf{k}$ is implicit in the $W_\alpha$ function, which takes the trace of occupied states. Because the matrices of $\boldsymbol{\mathcal{W}}$ are diagonal, and because the density functional terms are equal for all $\mathbf{k}$ by symmetry, we can write the above equation both more explicitly and more simply as
\begin{equation}
    E_\text{c}^\text{ACPT2} = \frac{\Omega}{8\pi^3} \int \dd[3]\mathbf{k}\,  \Theta(k_\text{F} - |\mathbf{k}|) \int_0^1 \dd\alpha \,2W_\alpha(W_0(k), W_0'(k), W_\infty, W_\infty')
\end{equation}
assuming dependence of $W_\alpha$ on only the four quantities $W_0, W_0', W_\infty, W_\infty'$, as is the case for ISI and modISI. The factor of 2 comes from the fact that there are 2 electrons per wave-vector $\mathbf{k}$ in the non-spin-polarized gas. All four terms are dependent on $k_\text{F}$ as well. The $W_\infty$ and $W_\infty'$ density functionals are evaluated using $n$ from eq~\ref{eq:rho_rs_kf} and $\nabla n=0$. $W_0(k)$ is simply one-half the exchange contribution to the eigenvalue at $k$, i.e.~\cite{martinElectronicStructureBasic2004}
\begin{align}
    W_0(k) &= -\frac{k_\text{F}}{2\pi}f_\text{x}\left(\frac{k}{k_\text{F}}\right) \\
    f_\text{x}(x) &= 1+\frac{1-x^2}{2x}\ln\left|\frac{1+x}{1-x}\right|
\end{align}
Returning to the correlation energy integral, we can divide by the number of electrons $N=\Omega k_\text{F}^3/(3\pi^2)$, convert to radial coordinates, and scale $k$ into $[0,1]$ to obtain
\begin{equation}
    \epsilon_\text{c}^\text{ACPT2} = \frac{E_\text{c}^\text{ACPT2}}{N} = 3\int_0^1 k^2 \dd k\, W_\alpha(W_0(k_\text{F}k), W_0'(k_\text{F}k), W_\infty, W_\infty') \label{eq:ec_acpt2}
\end{equation}

\subsection{Grids for Numerical Integration}

In this section, we explain how eqs~\ref{eq:w0p_ueg} and~\ref{eq:ikpq} are evaluated computationally. Note that all wave-vector units in this section are relative to the Fermi vector $k_\text{F}$ and therefore unitless. The restricted integration bounds in eq~\ref{eq:ikpq} are integrated numerically using Legendre-Gauss quadrature with $N_\text{sph}$ quadrature points. Note that it is important to scale the quadrature into the restricted integration bounds $[x_-(k,q), 1]$ and $[-1, x_+(p,q)]$, and not to integrate the Heaviside step functions numerically (even though these techniques are equivalent as $N_\text{sph}\rightarrow\infty$), because the discontinuity in the Heaviside step function causes slow convergence of the integral with respect to $N_\text{sph}$.

The integrals over the occupied wave-vectors $k$ and $p$ are performed using the following $N_\text{L}$-point quadrature
\begin{align}
    x_g &= \frac{g + 1/2}{N_\text{L}} \\
    k_g &= \frac{1 - \exp(-\alpha x_g)}{1 - \exp(-\alpha)} \\
    \dd k_g &= \frac{\alpha}{N_\text{L}} \frac{\exp(-\alpha x_g)}{1 - \exp(-\alpha)}
\end{align}
where $g={0,1,...,N_\text{L}-1}$ are the grid indexes. This quadrature essentially applies the midpoint rule to a uniform grid $x_g$, which is nonlinearly transformed to make the spacing between $k_g$ very dense near the Fermi level and sparser far away from the Fermi level. The reason for this is that the largest contributions to the correlation energy are near the Fermi level, and these contributions are very sensitive to the double-excitation gap, which in turn vanishes near the Fermi level.

To integrate the excitation wave-vector $q$, we use Treutler-Ahlrichs M4 integration grids~\cite{treutlerEfficientMolecularNumerical1995} with $N_\text{U}$ points, as implemented in PySCF~\cite{sunRecentDevelopmentsPySCF2020}, linearly scaled to have a maximum value of $q_{N_\text{U}-1} = q_\text{max}$ (with 0-indexing of the grids). It is critical that $q_g$ be very dense near zero, corresponding to large-wavelength excitations near the Fermi level. These excitations make the largest contributions to the correlation energy and will not be integrated efficiently on a more uniform grid. Farther from the Fermi level, the integrand varies much more slowly with respect to $q$, and the integration in turn can be much coarser. Both these properties are satisfied by the M4 grids, since they were designed to integrate density functionals on atomic densities, which are very large and sharp near the core and more diffuse at the atomic tails.

The quality of the numerical integrations therefore depends on four quantities: $N_\text{sph}, N_\text{L}, N_\text{U},$ and $q_\text{max}$. In this work, we use $N_\text{sph}=16$, $N_\text{L}=N_\text{U}=1000$, and $q_\text{max}=40$. With these settings, we find that the exchange diagram contribution to the GL2 correlation energy (see eqs~\ref{eq:w0p_gl2}--\ref{eq:gl2_ex} below) is 0.024179181600, only $2\times 10^{-8}$ Ha off the exact result. In addition, in Table~\ref{tab:ueg_conv}, we report the errors of these settings compared to a higher-fidelity setting for each quantity. The errors are all quite small over a wide range of densities with $r_\text{s}\in[10^{-2},10^2]$, with absolute errors ranging from $10^{-8}$ to $10^{-5}$ Ha. Relative errors are also quite small, with the exception of the $\kappa$-MP2 energies for very small densities $r_\text{s}>10$, but these errors are only large because the total $\kappa$-MP2 correlation energy is unphysically small, c.f.\ Figure 2 in the main text. Table~\ref{tab:ueg_conv2} shows that the relative errors for $\kappa$-MP2 are also small in the more reasonable density range $r_\text{s}\in[0.1,10]$.

\begin{table}[]
    \centering
    \caption{Maximum errors in the UEG value of $\kappa$-MP2 and OSMI-ACPT2 correlation energies with the default settings compared to more robust settings, taken over a range of densities from $r_\text{s}=0.01$ to $r_\text{s}=100$. ``Abs.'' indicates absolute error, and ``Rel.'' indicates relative error. The energy units are Hartree.}
    \setlength{\tabcolsep}{12pt}
    \begin{tabular}{lrrrr}
        \hline
        Settings Change & $\kappa$-MP2 Abs. & $\kappa$-MP2 Rel. & OSMI-ACPT2 Abs. & OSMI-ACPT2 Rel. \\
        \hline
        $N_\text{sph}=48$ & $2\times 10^{-6}$ & $1\times 10^{-5}$ & $6\times 10^{-7}$ & $3\times 10^{-6}$ \\
        $N_\text{L}=4000$ & $2\times 10^{-5}$ & $3\times 10^{-5}$ & $4\times 10^{-8}$ & $2\times 10^{-7}$ \\
        $N_\text{U}=4000$ & $7\times 10^{-6}$ & $4\times 10^{-3}$ & $3\times 10^{-8}$ & $2\times 10^{-7}$ \\
        $q_\text{max}=80^*$ & $3\times 10^{-7}$ & $1\times 10^{-1}$ & $2\times 10^{-7}$ & $1\times 10^{-6}$ \\
        \hline
    \end{tabular}\\
    $^*$Changing $q_\text{max}$ while keeping $N_\text{U}$ the same simultaneously changes the density of the grid near the Fermi level as well as the maximum excitation wave vector considered. For a fair comparison, we construct one grid with $N_\text{U}=2000, q_\text{max}=80$, and then we report the energy difference between this grid and the same grid but with values of $q_g>40$ excluded. The latter grid is not the exact same as that used for the results in this work. However, the comparison does quantify the error due to setting the maximum wave-vector to $q_\text{max}=40$ compared to raising to $q_\text{max}=80$.
    \label{tab:ueg_conv}
\end{table}

\begin{table}[]
    \centering
    \caption{Same as Table~\ref{tab:ueg_conv}, but only covering densities from $r_\text{s}=0.1$ to $r_\text{s}=10$.}
    \setlength{\tabcolsep}{12pt}
    \begin{tabular}{lrrrr}
        \hline
        Settings Change & $\kappa$-MP2 Abs.$^a$ & $\kappa$-MP2 Rel.$^b$ & OSMI-ACPT2 Abs.$^c$ & OSMI-ACPT2 Rel.$^d$ \\
        \hline
        $N_\text{sph}=48$ & $2\times 10^{-6}$ & $1\times 10^{-5}$ & $3\times 10^{-7}$ & $2\times 10^{-6}$ \\
        $N_\text{L}=4000$ & $8\times 10^{-8}$ & $4\times 10^{-7}$ & $1\times 10^{-8}$ & $1\times 10^{-7}$ \\
        $N_\text{U}=4000$ & $1\times 10^{-7}$ & $2\times 10^{-5}$ & $1\times 10^{-8}$ & $2\times 10^{-7}$ \\
        $q_\text{max}=80$ & $3\times 10^{-7}$ & $3\times 10^{-4}$ & $2\times 10^{-7}$ & $1\times 10^{-6}$ \\
        \hline
    \end{tabular}\\
    \label{tab:ueg_conv2}
\end{table}

\subsection{Divergence of GL2 for the Uniform Electron Gas}

For GL2, the effective potential of the zeroth-order Hamiltonian is a constant everywhere and can be neglected, so the eigenvalues are given by $E(k)=k^2/2$. Starting with eq~\ref{eq:w0p_ueg}, we can simplify the expression for the correlation energy matrix
\begin{align}
    W_0'(k) &= -\frac{1}{\pi^2} \int_0^1 p^2\dd p \int_0^\infty q^2\dd q \, I(k/k_\text{F},p,q) \label{eq:w0p_gl2} \\
    I(k,p,q) &= \int_{x_-(k,q)}^{1} \dd x \int_{-1}^{x_+(p,q)} \dd y \frac{2q^{-4} - q^{-2}\tilde{Q}^{-2}}{\Delta(k,p,q,x,y)} \\
    \Delta(k,p,q,x,y) &= q^2 + (kx-py)q
\end{align}

The exchange term of the correlation energy (the $q^{-2}\tilde{Q}^{-2}$ term) actually converges and has the value
\begin{equation}
    \epsilon_\text{c}^\text{exch} = \frac{1}{6} \ln 2 - \frac{3}{4\pi^2}\zeta(3) \approx 0.024179158918 \label{eq:gl2_ex}
\end{equation}
where $\zeta(s)$ is the Riemann zeta function. Therefore, we focus on the direct term
\begin{equation}
    W_\text{direct}(k_\text{F}k) = -\frac{2}{\pi^2} \int_0^1 p^2\dd p \int_0^\infty q^2\dd q \, \int_{x_-(k,q)}^{1} \dd x \int_{-1}^{x_+(p,q)} \dd y \frac{q^{-4}}{q^2+(kx-py)q}
\end{equation}
We can get the correlation energy by replacing $W_\alpha$ with $\alpha W_0'$ in eq~\ref{eq:ec_acpt2}:
\begin{equation}
    \epsilon_\text{c}^\text{MP2,direct} = -\frac{3}{\pi^2} \int_0^1 k^2\dd k \int_0^1 p^2\dd p \int_0^\infty q^2\dd q \, \int_{x_-(k,q)}^{1} \dd x \int_{-1}^{x_+(p,q)} \dd y \frac{q^{-4}}{q^2+(kx-py)q}
\end{equation}
Note that this equation is entirely independent of $k_\text{F}$ and can also be written as
\begin{align}
    \epsilon_\text{c}^\text{MP2,direct} &= -\frac{3}{\pi^2} \int_0^\infty q^{-2} \dd q\, f(q) \label{eq:mp2_from_fq} \\
    f(q) &= \int_0^1 k^2\dd k \int_0^1 p^2\dd p \int_{x_-(k,q)}^{1} \dd x \int_{-1}^{x_+(p,q)} \dd y \frac{1}{q^2+(2kx-2py)q}
\end{align}

Infrared divergence occurs due to small-wavelength behavior of the integrand, so we are interested in the limit $q\rightarrow 0$. In this limit, the integration occurs over a thin shell of width $q\cos\theta$ at the Fermi surface, and the integral can be rewritten as
\begin{equation}
    f(q) = \int_0^1 dx \int_0^1 dy \int_{1-qx}^1 k^2\dd k \int_{1-qy}^1 p^2 \dd p \frac{1}{q^2+(kx+py)q}
\end{equation}
Approximating the integral to lowest order in $q$ gives
\begin{equation}
    f(q) = q \int_0^1 dx \int_0^1 dy \frac{xy}{x+y} = \frac{2q}{3}\left(1 - \ln 2\right) \label{eq:fq_low_q}
\end{equation}
Using eq~\ref{eq:fq_low_q} and eq~\ref{eq:mp2_from_fq}, we see that there is a logarithmic divergence in the GL2 energy as the lower integration bound on $q$ vanishes:
\begin{equation}
    \epsilon_\text{c}^\text{MP2,direct} \approx -\frac{2}{\pi^2} (1 - \ln 2) \int_{q_\text{c}}^C q^{-1} \dd q\, \propto \ln q_\text{c} + \dots
\end{equation}
so $\epsilon_\text{c}^\text{MP2,direct}\rightarrow-\infty$ when $q_\text{c}$ vanishes. The GL2 correlation energy of the uniform electron gas is therefore not finite.

\section{Proof of Orbital-Invariance of PT2 Correlation Energy Matrix}

Consider unitary transformations $\mathbf{U}^O$ and $\mathbf{U}^V$ of the occupied and virtual molecular orbitals,
\begin{align}
    \ket{\tilde{\phi}_i} &= \sum_j U^O_{ij} \ket{\phi_j} \\
    \ket{\tilde{\phi}_a} &= \sum_b U^V_{ab} \ket{\phi_b}
\end{align}
Because $\mathbf{U}^O$ and $\mathbf{U}^V$ are unitary,
\begin{align}
    \ket{\phi_i} &= \sum_j U^O_{ji} \ket{\tilde{\phi}_j} \\
    \ket{\phi_a} &= \sum_b U^V_{ba} \ket{\tilde{\phi}_b}
\end{align}
Note that in this section we will assume real orbitals and real $\mathbf{U}^O$ and $\mathbf{U}^V$ for simplicity. 
Suppose now that we compute eq 18 of the main text in the basis of $\{\ket{\tilde{\phi}_i}\}$ rather than $\{\ket{\phi_i}\}$. We denote this matrix $\mathbf{\widetilde{W}}_0'$. For $\mathbf{\widetilde{W}}_0'$ to be orbital-invariant, we require that
\begin{equation}
    \left(\mathbf{U}^O\right)^\top \mathbf{\widetilde{W}}_0' \mathbf{U}^O = \mathbf{W}_0' \label{eq:orb_invariance}
\end{equation}
To prove that eq~\ref{eq:orb_invariance} holds, we start by noting that the t2 amplitudes and perturbation matrix elements in the $\{\ket{\tilde{\phi}_i}\}$ basis (denoted by $\tilde{t}$ and $\tilde{v}$) can be expressed via basis transformation of the same quantities in the $\{\ket{\phi_i}\}$ basis:
\begin{align}
    \tilde{v}_{jk}^{ab} = \sum_{lmcd} U_{ac} U_{bd} U_{km} U_{jl} v_{lm}^{cd} \\
    \tilde{t}_{ik}^{ab} = \sum_{lmcd} U_{ac} U_{bd} U_{km} U_{il} t_{lm}^{cd}
\end{align}
where $v_{ij}^{ab}\equiv\mel{ij}{}{ab}$. Using the above two equations and the unitary property $\sum_k U_{ik}U_{kj}=\delta_{ij}$, we can write
\begin{align}
    \sum_{kab} \tilde{t}_{ik}^{ab} \tilde{v}_{jk}^{ab} = \sum_{kabln} U_{in} U_{jl} t_{nk}^{ab} v_{lk}^{ab} \label{eq:transformation}
\end{align}
Evaluated in the $\{\ket{\tilde{\phi}_i}\}$ basis, eq 18 of the main text is is
\begin{align}
    \left(\mathbf{\widetilde{W}}_0'\right)_{ij} = \sum_{kab} \left(\tilde{t}_{ik}^{ab} \tilde{v}_{jk}^{ab} + \tilde{t}_{jk}^{ab} \tilde{v}_{jk}^{ab}\right) \label{eq:w0p_tilde}
\end{align}
Plugging eq~\ref{eq:transformation} into eq~\ref{eq:w0p_tilde} gives
\begin{equation}
    \left(\mathbf{\widetilde{W}}_0'\right)_{ij} = \sum_{kabln} U_{in} U_{jl} (\mathbf{W}_0')_{nl}
\end{equation}
or
\begin{equation}
    \mathbf{\widetilde{W}}_0' = \mathbf{U}^O \mathbf{W}_0' \left(\mathbf{U}^O\right)^\top
\end{equation}
Multiplying both sides by $\left(\mathbf{U}^O\right)^\top$ on the left and $\mathbf{U}^O$ on the right gives
\begin{equation}
    \left(\mathbf{U}^O\right)^\top \mathbf{\widetilde{W}}_0' \mathbf{U}^O = \mathbf{W}_0'
\end{equation}
as expected.

\section{An Illustration of Orbital Invariance}\label{sec:res_orb_inv}

Compared to size-consistency, the invariance of a quantum chemistry method to unitary transformations (rotations) of the occupied or virtual orbitals is a more difficult property to demonstrate explicitly. To do so, we compare the orbital-invariant OSMI method to the non-orbital-invariant modification OSVI introduced in the main text for cases where symmetrically distinct occupied orbitals cross each other in energy, resulting in the non-orbital-invariant method's energy becoming non-unique.

\begin{figure}
    \centering
    \includegraphics[width=0.4\linewidth]{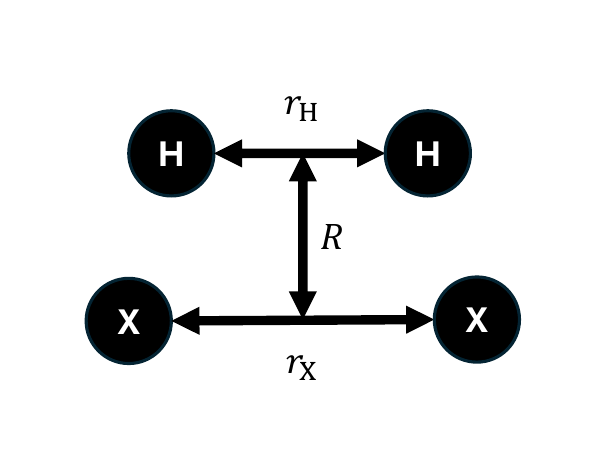}
    \caption{Molecular structure of the systems used to create numerically degenerate orbitals for tests of orbital-invariance. The circles are atoms, X is a changeable element, and the bonds of the two dimers are centered at the same point.}
    \label{fig:orb_inv}
\end{figure}

We start by designing non-spin-polarized molecular systems for which the two highest-energy occupied spatial orbitals are numerically degenerate but not related by symmetry. We do this by minimizing the difference between these two orbital eigenvalues as a function of $r_\text{H}$ in the system depicted in Figure~\ref{fig:orb_inv}. We found the bond length at which the two highest occupied orbitals are degenerate by minimizing the difference between their orbital energies as a function of the bond length using scipy.optimize.minimize. The other parameters of the molecular geometry were chosen by hand to make it easier to create degenerate orbitals.

Next, we redefine the two highest occupied orbitals as
\begin{align}
    \ket{\phi_{N_\text{occ}-1}} &\gets \frac{1}{\sqrt{2}} \left(\ket{\phi_{N_\text{occ}-1}} + \ket{\phi_{N_\text{occ}}}\right) \\
    \ket{\phi_{N_\text{occ}}} &\gets \frac{1}{\sqrt{2}} \left(\ket{\phi_{N_\text{occ}-1}} - \ket{\phi_{N_\text{occ}}}\right)
\end{align}
This rotation of the orbitals does not change the DFT energy or HF energy. In addition, the mean-field Hamiltonian is still diagonal in the new molecular orbital basis, so we can still evaluate the GL2 amplitudes using eqs 5 and 6 in the main text. We define the total energy evaluated on the original orbitals as $E_0$, the total energy evaluated on the rotated orbitals as $E_0^\text{rot}$, and
\begin{equation}
    \Delta E_\text{rot} = E_0 - E_0^\text{rot} \label{eq:erot}
\end{equation}
We expect that $\Delta E_\text{rot} = 0$ for an orbital-invariant method. As shown in Table~\ref{tab:orb_inv} for two systems of the structure specified in Figure~\ref{fig:orb_inv}, $\Delta E_\text{rot}$ does in fact vanish (to reasonable numerical precision) for the OSMI method, but is finite for OSVI. This demonstrates empirically that OSMI is orbital-invariant, while OSVI is not.

We note that while this somewhat pathological model demonstrates the effect of orbital-invariance most clearly, OSMI and OSVI give significantly different energy predictions even for very simple systems. For example, in the def2-TZVP basis, the OSMI and OSVI methods with the modISI model give correlation energies for the Ar atom of -0.3148 Ha and -0.3172, respectively, which differ by 0.0024 Ha.

\begin{table}[]
    \caption{Effect of rotating numerically degenerate orbitals for the orbital-invariant (OSMI) and non-orbital-invariant (OSVI) models. See Figure~\ref{fig:orb_inv} for the definitions of the structure parameters X, $r_\text{H}$, $r_\text{X}$, and $R$, and see eq~\ref{eq:erot} for the definition of $\Delta E_\text{rot}$. Distance units are in \r{A}, and energy units are in Ha. These results were obtained using a def2-TZVP basis~\cite{weigendBalancedBasisSets2005} without density fitting.}
    \centering
    \setlength{\tabcolsep}{12pt}
    \begin{tabular}{lrrrrr}
        \hline
        X & $r_\text{H}$ & $r_\text{X}$ & $R$ & $\Delta E_\text{rot}$ (OSMI) & $\Delta E_\text{rot}$ (OSVI) \\
        \hline
        F & 0.79260661 & 1.40 & 2.0 & $<10^{-11}$ & $-1.2 \times 10^{-4}$ \\
        N & 0.77787118 & 1.10 & 10.0 & $<10^{-11}$ & $1.9 \times 10^{-5}$ \\
        \hline
    \end{tabular}
    \label{tab:orb_inv}
\end{table}

\section{ISI vs modISI}

\begin{figure}
    \centering
    \includegraphics[scale=1.0]{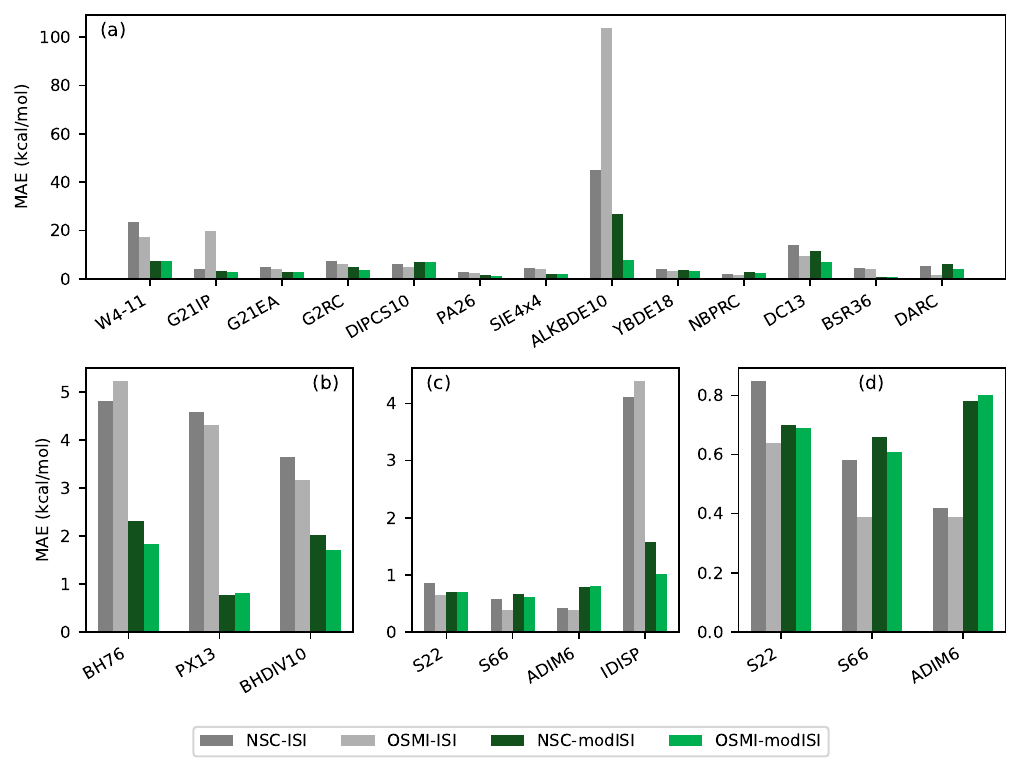}
    \caption{Mean absolute errors (MAEs) of the ISI and modISI adiabatic connection functionals using the traditional non-size-consistent (NSC) approach and the OSMI approach developed in this work. The subsets are separated based on type of property: (a) reaction energies, ionization potentials, and electron affinities, (b) barrier heights, (c) noncovalent interactions, and (d) noncovalent interactions excluding IDISP, to make the energy scale more readable.}
    \label{fig:isi_vs_modisi}
\end{figure}

As mentioned in the main text, we chose the modISI interpolation of this work in place of the original ISI model of Seidl, Perdew, and Kurth~\cite{seidlSimulationAllOrderDensityFunctional2000} because ISI yielded poor results for some thermochemical benchmark sets. The modISI model does not suffer from this issue, though it has the drawback that (to the best of our knowledge) there is not an analytical form for the integral over $\alpha$.

This problem is illustrated in fig~\ref{fig:isi_vs_modisi}. Most notably, the MAE is more than twice as large with ISI than with modISI for the W4-11 atomization energies, and the ISI MAE is 44 kcal/mol with the NSC approach and 104 kcal/mol with the OSMI approach for the ALKBDE10 dataset. ISI is also much worse than modISI for barrier heights, with larger MAEs by 1.5-3.5 kcal/mol. We are not sure why the errors with ISI are so large for some subsets. It could be an interesting direction to explore in future work, since ISI is more accurate than modISI for some dispersion datasets like S66 and ADIM6.

\section{MP2 vs RMP2}

\begin{figure}
    \centering
    \includegraphics[scale=1.0]{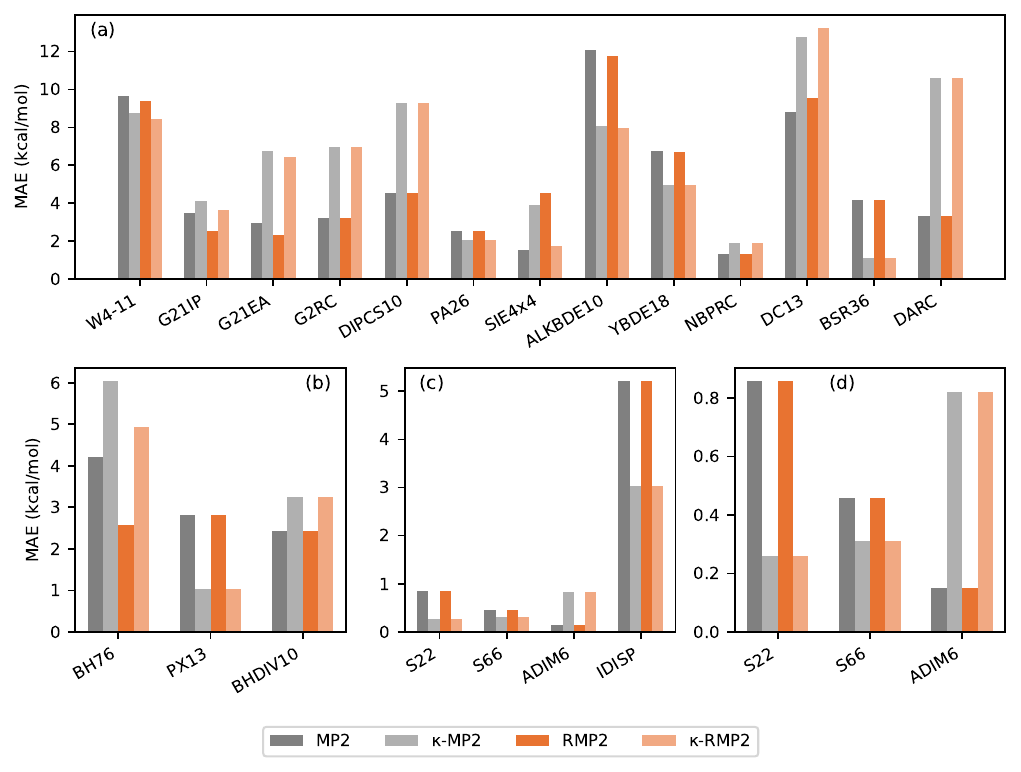}
    \caption{Mean absolute errors (MAEs) of the MP2 and RMP2 methods. They are identical for many subsets because MP2 and RMP2 are identical for closed-shell systems. The subsets are separated based on type of property: (a) reaction energies, ionization potentials, and electron affinities, (b) barrier heights, (c) noncovalent interactions, and (d) noncovalent interactions excluding IDISP, to make the energy scale more readable.}
    \label{fig:mp2_vs_rmp2}
\end{figure}

Spin-contamination is known to degrade the accuracy of MP2 for some open-shell systems~\cite{sheeRegularizedSecondOrderMoller2021a}, which is why we use RMP2 in the main text. A comparison of MP2 and RMP2 is provided in Figure~\ref{fig:mp2_vs_rmp2}. Most datasets are similar or identical with MP2 and RMP2, but MP2 is significantly less accurate than RMP2 for the BH76 barrier heights database. RMP2 is never significantly worse than MP2 for these systems, with the exception of SIE4x4. For this subset, MP2 outperforms RMP2, but $\kappa$-RMP2 still outperforms $\kappa$-MP2.

\section{GMTKN55 Benchmarks with def2 basis sets}

For completeness, we also performed the GMTKN55 benchmarks from the main text with T,Q extrapolation to the complete basis set using the def2-XZVPPD basis sets (X=T,Q). The results are mostly similar to the aug-CC-PVXZ results, but for a few subsets, such as ALKBDE10 and G21EA, there are significant differences ($>1$ kcal/mol difference for the MAE for some methods).

\begin{figure}
    \centering
    \includegraphics[scale=1.0]{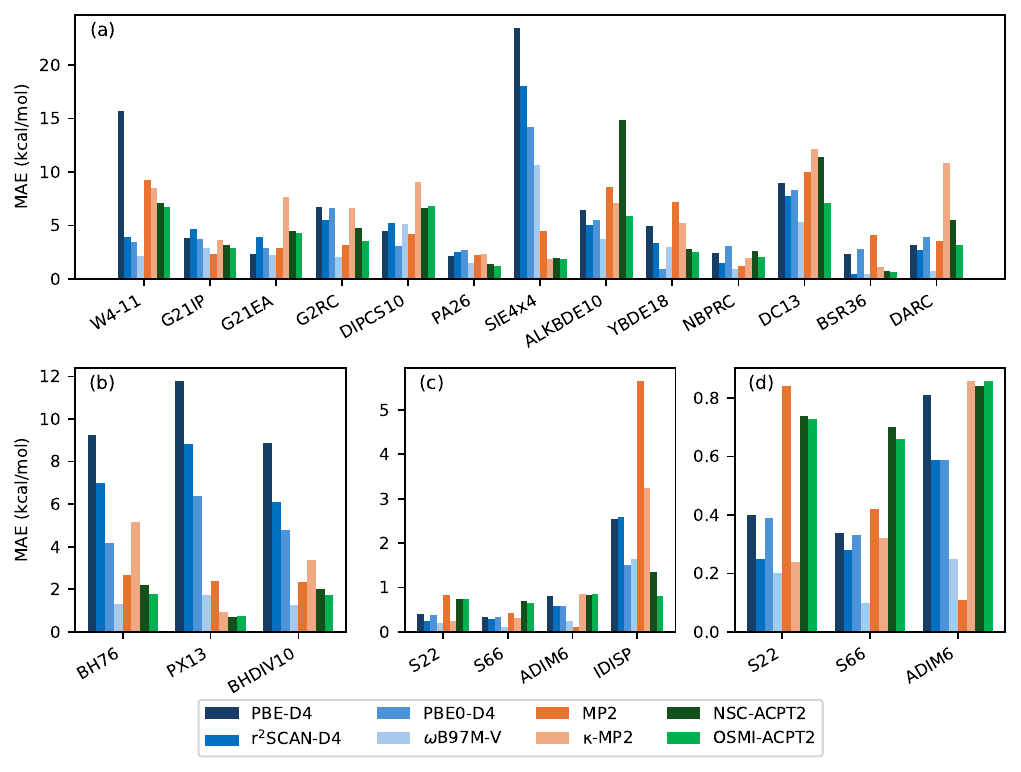}
    \caption{Same as Figure 3 of the main text, but with results computed using the def2-XZVPPD basis sets instead of aug-cc-pVXZ.}
    \label{fig:gmtkn55_si}
\end{figure}

\section{Size-consistency effects of Core Correlation for Bond Dissociation}

\begin{figure}
    \centering
    \includegraphics[scale=1.0]{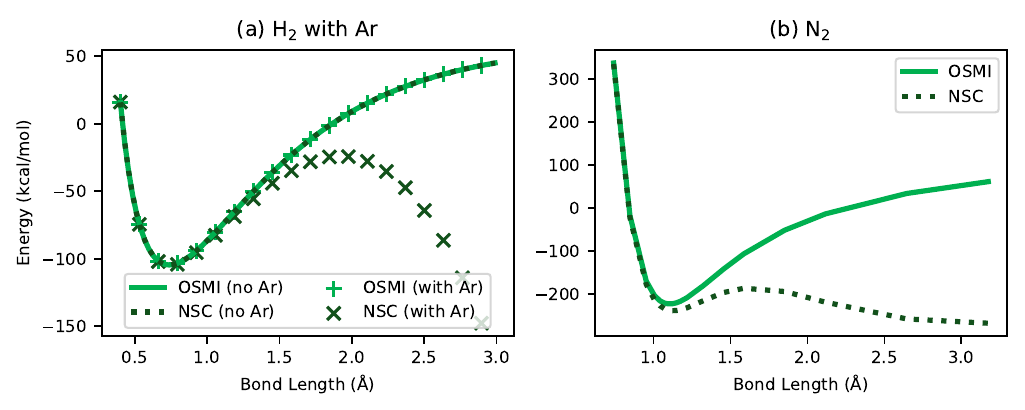}
    \caption{Dissociation curves of the hydrogen molecule with and without an argon ``spectator atom'' 100 \r{A} away (left) and of the nitrogen molecule (right).  ``NSC'' is the non-size-consistent ISI model, and ``SC'' is the size-consistent modISI model. The hydrogen dissociation curves with and without argon match for the size-consistent model but not the non-size-consistent model. Energies are referenced to the spin-unrestricted dissociation limit for each method. The aug-cc-pCVQZ basis was used to generate the dissociation curves, and the core electrons were not frozen.}
    \label{fig:sc}
\end{figure}

In the main text, we mentioned that correlation effects of core electrons strongly affect non-size-consistent adiabatic connection methods (degrading the accuracy of the method compared to when the core electrons are frozen), and we illustrated this by comparing the W4-11 atomization energies with and without frozen core electrons (main text Table 1). The effects of core correlation are even more extreme when static correlation comes into play, such as in the bond dissociation curves of Figure 1 in the main text. This is shown in Figure~\ref{fig:sc}, which contains the same dissociation curves as Figure 1 computed with the aug-cc-pCVQZ basis. The divergence of the NSC model is even more extreme for the Ar spectator problem, as is its over-correlation of the dissociated \ce{N2} molecule.

\section{Alternative Visualization of \ce{He2+} Dissociation Curve}

\begin{figure}
    \centering
    \includegraphics[scale=1.0]{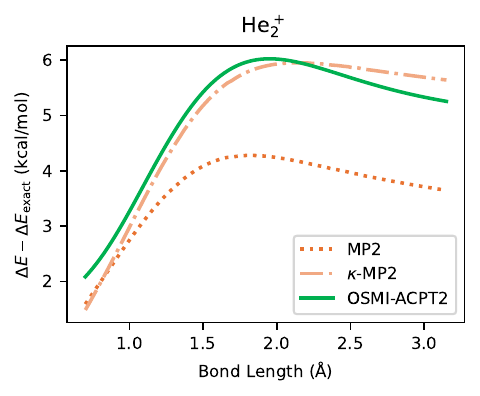}
    \caption{Dissociation curve of \ce{He2+} with different methods, minus the ``nearly exact'' values computed with CCSD(T). $\Delta E$ refers to the energy difference between the stretched bonded system and its open-shell dissociation limit for a given method.}
    \label{fig:he2p}
\end{figure}

Because MP2, $\kappa$-MP2, and OSMI-ACPT2 provide very similar dissociation curves for \ce{He2+}, we plot the difference between these dissociation curves and the reference dissociation curve in Figure~\ref{fig:he2p} for clearer visualization.

\bibliography{mybib}